\documentclass[prd,aps,11pt,superscriptaddress,nofootinbib,twoside,raggedbottom,amsmath]{revtex4-1}

\usepackage{xpatch}
\makeatletter
\patchcmd{\@ssect@ltx}
    {\addcontentsline{toc}{#1}{\protect\numberline{}#8}}
    {}
    {}
    {}
\makeatother

\usepackage{hyperref}
\hypersetup{
    colorlinks = true,
    citecolor  = red,
	linkcolor  = blue
}

\usepackage[utf8]{inputenc}
\usepackage{graphicx} 
\usepackage[dvipsnames,table,xcdraw]{xcolor}
\usepackage{amsmath}
\usepackage{bbold}
\usepackage{cases}
\usepackage{mathrsfs}
\usepackage[version=4]{mhchem}
\usepackage{booktabs}

\newcommand{\be}{\begin{equation}}
\newcommand{\ee}{\end{equation}}
\newcommand{\beq}{\begin{eqnarray}}
\newcommand{\eeq}{\end{eqnarray}}
\newcommand{\bee}{\begin{eqnarray}}
\newcommand{\eee}{\end{eqnarray}}
\def\bit{\begin{itemize}}
\def\eit{\end{itemize}}
\def\ben{\begin{enumerate}}
\def\een{\end{enumerate}}

\newcommand{\Eq}[1]{Eq.~(\ref{#1})}

\makeatletter
\newcommand{\thickhline}{%
    \noalign {\ifnum 0=`}\fi \hrule height 1pt
    \futurelet \reserved@a \@xhline
}

\newcolumntype{?}{!{\vrule width 1pt}}

\makeatletter
\renewcommand*\env@matrix[1][\arraystretch]{%
  \edef\arraystretch{#1}%
  \hskip -\arraycolsep
  \let\@ifnextchar\new@ifnextchar
  \array{*\c@MaxMatrixCols c}}
\makeatother

\newcommand{\rvline}{\hspace*{-\arraycolsep}\vline\hspace*{-\arraycolsep}}

\newcommand{\OO}{\mathcal{O}}

\newcommand{\bk}{\mathbf{k}}

\newcommand{\GeV}{{\,\rm GeV}}

\newcommand{\meV}{{\,\rm meV}}
\newcommand{\MeV}{{\,\rm MeV}}

\newcommand{\T}{{\,\rm T}}

\DeclareMathAlphabet\mathbfcal{OMS}{cmsy}{b}{n} 

\def\vect#1{\boldsymbol{#1}}
\def\tens#1{\mathbf{#1}}

\graphicspath{{./figures/}}

\definecolor{cerulean}{rgb}{0., 0.52,0.65}

\newcommand\Tstrut{\rule{0pt}{3.5ex}}         
\newcommand\Bstrut{\rule[-2ex]{0pt}{0pt}}   

\begin{document}

\title{Detectability of Axion Dark Matter with Phonon Polaritons and Magnons}
\author{Andrea~Mitridate}
\affiliation{Walter Burke Institute for Theoretical Physics, California Institute of Technology, Pasadena, CA 91125, USA}

\author{Tanner~Trickle}
\affiliation{Department of Physics, University of California, Berkeley, CA 94720, USA}
\affiliation{Theoretical Physics Group, Lawrence Berkeley National Laboratory, Berkeley, CA 94720, USA}
\affiliation{Walter Burke Institute for Theoretical Physics, California Institute of Technology, Pasadena, CA 91125, USA}

\author{Zhengkang~Zhang}
\affiliation{Walter Burke Institute for Theoretical Physics, California Institute of Technology, Pasadena, CA 91125, USA}

\author{Kathryn~M.~Zurek}
\affiliation{Walter Burke Institute for Theoretical Physics, California Institute of Technology, Pasadena, CA 91125, USA}

\begin{abstract}
\vspace{10pt}
Collective excitations in condensed matter systems, such as phonons and magnons, have recently been proposed as novel detection channels for light dark matter. We show that excitation of {\em i)} optical phonon polaritons in polar materials in an $\OO$(1\,T) magnetic field (via the axion-photon coupling), and {\em ii)} gapped magnons in magnetically ordered materials (via the axion wind coupling to the electron spin), can cover the difficult-to-reach $\OO(1$\,-\,$100)\,$meV mass window of QCD axion dark matter with less than a kilogram-year exposure. Finding materials with a large number of optical phonon or magnon modes that can couple to the axion field is crucial, suggesting a program to search for a range of materials with different resonant energies and excitation selection rules; we outline the rules and discuss a few candidate targets, leaving a more exhaustive search for future work. Ongoing development of single photon, phonon and magnon detectors will provide the key for experimentally realizing the ideas presented here.
\end{abstract}

\maketitle
\newpage
\tableofcontents
\thispagestyle{empty}
\newpage


\section{Introduction}

The QCD axion \cite{Weinberg1975a,Peccei1977a,Peccei1977b,Wilczek1978a} remains one of the best-motivated and predictive models of dark matter (DM). The search for the axion has a decades long history, and is still ongoing.  At the moment, only the Axion Dark Matter Experiment (ADMX)~\cite{Asztalos2010a,Du2018a} has sensitivity to the QCD axion in a narrow mass range around 2-3\,$\mu$eV. The HAYSTAC experiment is seeking to extend these results to higher frequencies~\cite{Zhong2018a}. The ABRACADABRA~\cite{Ouellet:2018beu} and CASPEr \cite{Garcon:2019inh} experiments have also recently achieved their first limits for very light masses (though with sensitivity still far above that needed to reach the QCD axion). The CERN Axion Solar Telescope (CAST)~\cite{Anastassopoulos2017a} is searching for axions emitted by the Sun, and can constrain the QCD axion for masses above $\sim 1$\,eV.  Many more experiments plan to join this search.  These include the MAgnetized Disk and Mirror Axion eXperiment (MADMAX)~\cite{TheMADMAXWorkingGroup:2016hpc,Brun2019a}, which uses a layered dielectric in an external magnetic field, and the QUaere AXion (QUAX) experiment~\cite{Ruoso:2015ytk,Barbieri2017a,Crescini:2018qrz}, which searches for axion-induced classical spin waves inside a magnetic target. See also Ref.~\cite{DeRocco:2018jwe,Obata:2018vvr,Liu:2018icu,Flower:2018qgb,Nagano:2019rbw,Lawson:2019brd,Berlin:2019ahk,Lasenby:2019prg,Lasenby:2019hfz} for recent axion DM search proposals.

The QCD axion mass window $m_a\sim\OO(1$\,-\,$100)$\,meV remains, however, unconstrained. The current best limits are provided by CAST, but this could be outperformed in the future by dish antennas~\cite{Horns2013a} or multilayer films~\cite{Baryakhtar:2018doz} (both of which are related to MADMAX in concept but can reach higher axion masses).  These are limited by current single photon detection technology, which is rapidly improving. Recently the use of axionic topological antiferromagnets has been proposed to detect axions in this region~\cite{Marsh:2018dlj}, although such materials have not been fabricated yet in the lab, and even then, this proposal is limited to $m_a \lesssim 10\meV$.

\begin{figure}[t]
\centering
\includegraphics[width=0.95\textwidth]{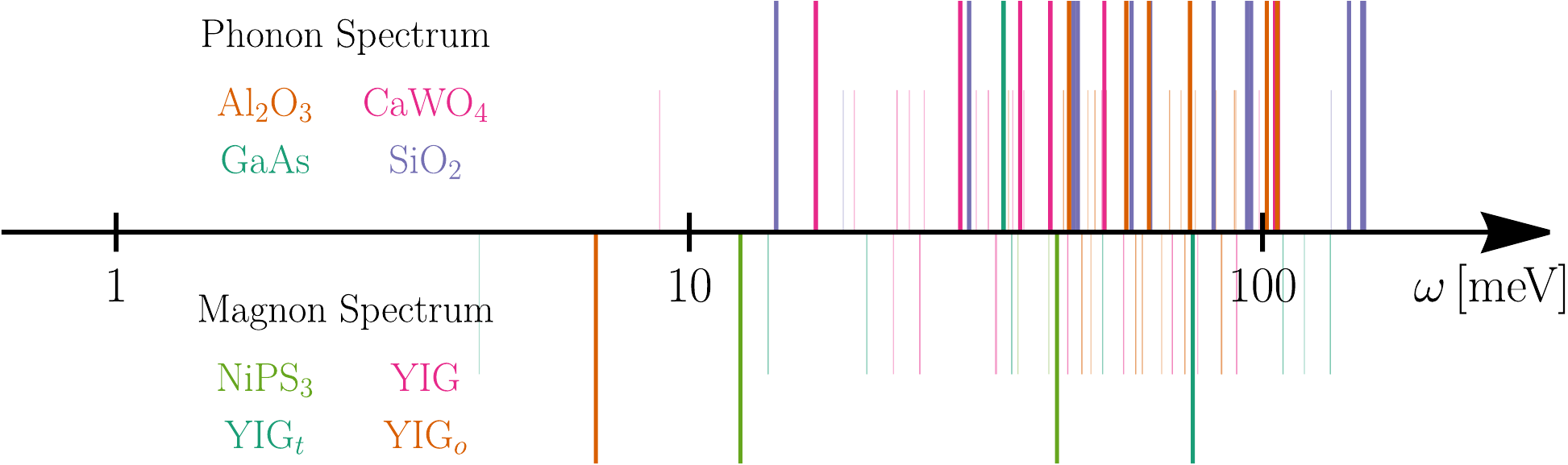}
\caption{
Spectra of gapped phonon polaritons and magnons at zero momentum for several representative targets considered in this work. These collective excitations have typical energies of $\OO(1$\,-\,$100)$\,meV, and can be utilized to search for axion DM in the mass window $m_a\sim\OO(1$\,-\,$100)$\,meV. Longer lines with darker colors correspond to the resonances in Figs.~\ref{fig:phoreach1}, \ref{fig:phoreach2} and \ref{fig:magreach}, while the shorter ones with lighter colors represent modes with suppressed couplings to axion DM due to selection rules. 
\label{fig:spectrum}}
\end{figure}

Collective excitations, such as phonons and magnons, have resonance energies in the $\OO(1$\,-\,$100)$~meV range, as shown in Fig.~\ref{fig:spectrum}.  They have been proposed as an excellent way to detect light dark matter through scattering (if the dark matter is heavier than a keV) or absorption (if the dark matter is in the $\OO(1$\,-\,$100)$~meV mass window)~\cite{Hochberg:2016ajh,Schutz:2016tid,Knapen:2016cue,Hochberg:2016sqx,Knapen:2017ekk,Griffin:2018bjn,Acanfora:2019con,Trickle:2019ovy,Caputo:2019cyg,Trickle:2019nya,Griffin:2019mvc,Baym:2020uos}. While previous work has shown the reach to dark photon absorption, an open question is whether phonon and magnon excitations possess a sufficiently strong coupling to reach the QCD axion. 

In this paper, we investigate axion absorption onto phonons and magnons, and demonstrate the potential of these processes to cover the $m_a\sim\OO(1$\,-\,$100)$\,meV QCD axion mass window. The particle-level axion interactions of interest are: 
\begin{equation}
{\cal L} = -\frac{1}{4} g_{a\gamma\gamma} a\,F_{\mu\nu} \tilde F^{\mu\nu} 
+\sum_{f=e,p,n} \frac{g_{aff}}{2m_f} (\partial_\mu a) (\bar f \gamma^\mu \gamma^5 f)
-\sum_{f=p,n} \frac{g_{af\gamma}}{4} a F_{\mu\nu} (\bar f i\sigma^{\mu\nu} \gamma^5 f )\,,
\label{eq:La}
\end{equation}
where the three terms are the axion's electromagnetic, wind and electric dipole moment (EDM) couplings, respectively. In the nonrelativistic limit, the effective interaction Hamiltonian is\footnote{The coupling to the axial current also generates a term proportional to $m_a \vect{s}_f\cdot\vect{v}_f$, we neglect this term since its coupling to collective spin excitations is suppressed compared to the one generated by the $\nabla a\cdot \vect{s}_f$ term.}
\begin{equation}
 \delta \hat{H} = -g_{a\gamma\gamma} \int d^3 x \, a \vect{E}\cdot\vect{B} 
-\sum_{f=e,p,n} \frac{g_{aff}}{m_f} \,\nabla a\cdot \vect{s}_f
- \sum_{f=p,n} g_{af\gamma} \,  a \,  \vect{E}\cdot\vect{s}_f\,.
\label{eq:non_rel_int_ham}
\end{equation}
These couplings can be further matched onto axion couplings to low energy degrees of freedom in a crystal. In particular, phonon excitation results from couplings to atomic displacements $\vect{u}_{lj} = \vect{x}_{lj} -\vect{x}_{lj}^0$, where $l$ labels the primitive cell, $j$ labels the atoms within each cell, and $\vect{x}_{lj}^0$ are the equilibrium positions, while magnons can be excited via couplings to the (effective) spins of magnetic ions $\vect{S}_{lj}$. An axion field oscillating with frequency $\omega=m_a$ and wavenumber $\vect{p} = m_a \vect{v}_a$ is represented by 
\begin{equation}
a(\vect{x},t) = a_0\, \cos{\left( \vect{p}\cdot\vect{x} - \omega t \right)} \,,
\label{eq:axt}
\end{equation}
where the field amplitude is related to the energy density via $\rho_a = m_a^2 a_0^2/2$. The resulting effective Hamiltonian relevant for phonon and magnon production takes the general form
\begin{equation}
\delta \hat{H} = \delta \hat H_0 e^{-i\omega t}+\text{c.c.} \,,\qquad
\delta \hat H_0 = 
\begin{cases}
\sum\limits_{lj} e^{i\vect{p}\cdot\vect{x}_{lj}^0}\, \vect{f}_j \cdot \vect{u}_{lj} & \;\Rightarrow \;\;\; \text{phonons}\,, \\
\sum\limits_{lj} e^{i\vect{p}\cdot\vect{x}_{lj}^0}\, \vect{f}_j \cdot \vect{S}_{lj} & \;\Rightarrow \;\;\; \text{magnons}\,,
\end{cases}
\label{eq:fj}
\end{equation}
with the effective couplings, $\vect{f}_j$, proportional to $a_0$ and the relevant axion coupling. While our focus here is axion DM, the same equations hold for general field-like DM candidates.

In Sec.~\ref{sec:formalism}, we derive rate formulae for single phonon and magnon excitations starting from the general form of couplings in Eq.~\eqref{eq:fj}. In the case of phonon excitation, the true energy eigenmodes in a polar crystal, at the low momentum transfers relevant for dark matter absorption, are phonon polaritons due to the mixing between the photon and phonons. We take this mixing into account while still often referring to the gapped polaritons as phonons since their phonon components are much larger. The final results for phonon and magnon excitation rates are Eqs.~\eqref{eq:Rpho} and \eqref{eq:Rmag}. Depending on the couplings $\vect{f}_j$ and symmetries of the target system, it often happens that excitation of some of the phonon or magnon modes is suppressed, reducing the sensitivity to DM. We discuss this problem and possible ways to alleviate it in Sec.~\ref{sec:selection}.

Then it remains to determine the effective couplings $\vect{f}_j$ in terms of particle physics parameters -- $g_{a\gamma\gamma}$, $g_{aff}$, etc.\ -- in the case of axion DM. The effective couplings can receive multiple contributions, some of which rely on the presence of an external field. We discuss the various possibilities for axion-induced single phonon or magnon production in Sec.~\ref{sec:couplings}. Among them, two are particularly promising: the coupling of the gradient of the axion field to the electron spin, $g_{aee}$, allows for magnon excitation, while the axion-induced electric field in the presence of an external magnetic field, due to the axion-photon coupling $g_{a\gamma\gamma}$, can excite phonon polaritons. These processes are summarized in Table~\ref{tab:chan}. We present our numerical results for the projected reach via these processes in Sec.~\ref{sec:sensitivity}. We find that, when the axion mass is well-matched to phonon polariton or magnon resonances in the target material, the QCD axion can be easily within reach. The sensitivity is inherently narrow-band for any specific target material, with the axion masses covered limited by the resonance widths. However, combining the reach of a set of judiciously chosen materials with different phonon and magnon frequencies can offer a broader coverage. Finally, we conclude in Sec.~\ref{sec:conclusions} and discuss future interdisciplinary work needed to better understand and realize the potential of the ideas presented in this work.

\begin{table}[t]
\begin{tabular}{c|c|c|c}
\rowcolor[HTML]{EFEFEF} 
Process                       & \;Fundamental interaction\; & Effective coupling in Eq.~\eqref{eq:fj}                                                                         & \;Rate formula                         \Tstrut\Bstrut       \\ \hline\hline
Axion $+$ B field $\to$ phonon \;  & $a\vect{E}\cdot\vect{B}$ & $\vect{f}_j = \frac{1}{\sqrt{2}}\, g_{a\gamma\gamma} \frac{e \sqrt{\rho_a}}{m_a}\, \vect{B} \cdot \vect{\varepsilon}^{-1}_\infty \cdot \tens{Z}_j^*$ & Eq.~\eqref{eq:Rpho}  \Tstrut\Bstrut\\ \hline
Axion $\to$ magnon             & $\nabla a\cdot\vect{s}_e$ & \;$\vect{f}_j =-\frac{i}{\sqrt{2}}\, g_{aee} \,(g_j-1) \,\frac{\sqrt{\rho_a}}{m_e} \, \vect{v}_a$   & Eq.~\eqref{eq:Rmag} \Tstrut\Bstrut \\ \hline
\end{tabular}
\caption{\label{tab:chan} Summary of the potentially detectable channels identified in section \ref{sec:couplings}. The axion field $a$ is given by Eq.~\eqref{eq:axt}, $\rho_a$ is its energy density, and $\vect{v}_a$ is its velocity. The axion couplings $g_{a\gamma\gamma}$ and $g_{aee}$ are defined in Eqs.~\eqref{eq:La} and \eqref{eq:non_rel_int_ham}, and given by Eqs.~\eqref{eq:gayy} and \eqref{eq:gaff} for the QCD axion. $\varepsilon_\infty$ is the high-frequency dielectric constant due to electronic screening, $\tens{Z}_j^*$ is the Born effective charge tensor of the ion, and $g_j$ is the Land\'e $g$-factor. \tt{$\varepsilon$ here.}
}
\end{table}


\section{General formalism for absorption rate calculations}
\label{sec:formalism}

In this section, we adapt the DM scattering calculations in Refs.~\cite{Trickle:2019ovy,Trickle:2019nya} to the present case of bosonic DM absorption. Unlike the scattering case, light bosonic DM (denoted by $a$ in what follows) should be treated as a classical field. Within the coherence time $\tau_a= (m_a v_a^2)^{-1} \sim 10^{-7}\,\text{s} \,(10\,\text{meV}/m_a)$, its effect can be modeled as a harmonic perturbation on the target system as in Eq.~\eqref{eq:fj}. In this work, we focus on configurations with no external AC electromagnetic fields, so that $\omega=m_a$. An AC external field with frequency $\omega_e$ would generate perturbations with $\omega = |m_a\pm \omega_e|$, for which the calculations in this section also apply.

Phonons and magnons arise from quantizing crystal lattice degrees of freedom, displacements $\vect{u}_{lj}$ and effective spins $\vect{S}_{lj}$ respectively, which DM can couple to, as mentioned in the Introduction --- see Eq.~\eqref{eq:fj}. The effective couplings $\vect{f}_j$ depend on the atom/ion types, hence the subscript $j$. We will keep $\vect{f}_j$ general in this section, and derive their expressions for the case of axion DM in Sec.~\ref{sec:couplings}.

We assume the target system is prepared in its ground state $|0\rangle$ at zero temperature. The transition rate from standard time-dependent perturbation theory reads
\begin{equation}
\Gamma = \sum_f \bigl| \langle f | \hat{\delta H_0} | 0 \rangle \bigr|^2 \,2\pi\,\delta(\omega-\omega_f)\,.
\label{eq:Gamma_delta}
\end{equation}
Strictly speaking, since phonons and magnons are unstable particles, the sum over final states $f$ should include multi-particle states resulting from their decays. In practice, however, when $\omega$ is close to a phonon/magnon resonance, we can simply smear the delta function to the Breit-Wigner function and sum over single phonon/magnon states:\footnote{In deriving Eq.~\eqref{eq:Gamma} we have assumed the observation time $t\gtrsim\gamma_{\nu,\vect{k}}^{-1}$, for which the transition rate $\Gamma$ is time-independent. We also assume that the line width of the axion, $\Delta \omega \sim m_a v_a^2 \sim 10^{-8}\,\text{eV}\,(m_a/10\,\text{meV})$ is smaller than excitation linewidth, $\gamma_{\nu, \vect{k}}$, which is true as long as $\gamma_{\nu, \vect{k}}$ is greater than $\sim 10^{-6}$ times the resonance frequency.}  
\begin{equation}
\Gamma = \sum_{\nu,\vect{k}} \bigl| \langle \nu,\vect{k}\, | \hat{\delta  H_0} | 0 \rangle \bigr|^2 \,\frac{4\omega\,\omega_{\nu,\vect{k}}\gamma_{\nu,\vect{k}}}{(\omega^2-\omega_{\nu,\vect{k}}^2)^2+(\omega\gamma_{\nu,\vect{k}})^2}\,,
\label{eq:Gamma}
\end{equation}
where $|\nu,\vect{k}\rangle$ is the single phonon/magnon state on branch $\nu$ with momentum $\vect{k}$, and $\gamma_{\nu,\vect{k}}$ is its decay width. Away from a resonance, the lineshape deviates from Breit-Wigner, and depends on the details of phonon/magnon interactions. Since we are interested in sub-eV DM candidates, the momentum deposited is limited to $m_a v_a \lesssim$ meV --- the DM field drives phonon/magnon modes close to the center of the first Brillouin zone (1BZ). Finally, averaging over the DM velocity distribution $f(\vect{v})$, we obtain the expected total rate:
\begin{equation}
\langle\Gamma\rangle = \int d^3 v\, f(\vect{v})\, \Gamma(\vect{v})\,.
\label{eq:GammaAve}
\end{equation}
We take $f(\vect{v})$ to be a boosted Maxwell-Boltzmann distribution,
\beq
f_\chi^\text{MB}(\vect{v}) &=& \frac{1}{N_0} e^{-(\vect{v}+\vect{v}_\text{e})^2/v_0^2} \,\Theta \bigl(v_\text{esc}-|\vect{v}+\vect{v}_\text{e}|\bigr) \,,\\
N_0 &=& \pi^{3/2} v_0^2 \left[ v_0\,\text{erf} \bigl(v_\text{esc}/v_0\bigr) -\frac{2 \, v_\text{esc}}{\sqrt{\pi}}\, \exp\bigl(-v_\text{esc}^2/v_0^2\bigr)\right] .
\eeq
with parameters $v_0 = 230$ km$/$s, $v_\text{e} = 240$ km$/$s, $v_\text{esc} = 600$ km$/$s. The local axion DM density $\rho_a$, which enters the effective couplings $\vect{f}_j$ (see Table~\ref{tab:chan}), is assumed to be $0.3$ GeV$/$cm$^3$.

In the following subsections, we derive the rate formulae for single phonon and magnon excitations, respectively. For easy comparison, we present both derivations in as similar ways as possible.


\subsection{Phonon excitations}
\label{sec:formalism-pho}

We begin by calculating the absorption rate from couplings to phonons. The target Hamiltonian results from expanding the potential energy of the crystal around equilibrium positions of atoms:
\begin{equation}
\hat{H} =  \sum_{lj} \frac{\vect{p}_{lj}^2}{2m_j} + \frac{1}{2} \sum_{ll'jj'} \vect{u}_{lj} \cdot \tens{V}_{ll'jj'} \cdot \vect{u}_{l'j'} + \OO\left( \vect{u}^3 \right)\,,
\end{equation}
where $\vect{p}_{lj}=m_j\dot{\vect{u}}_{lj}$, and the force constant matrices, $\tens{V}_{ll'jj'}$, can be calculated from {\it ab initio} density functional theory (DFT) methods~\cite{Martin2004a,Kresse1993a,Kresse1994a,Kresse1996a}.

To diagonalize the Hamiltonian, we expand the atomic displacements and their conjugate momenta in terms of canonical phonon modes:
\beq
\vect{u}_{lj} &=&  \sum_{\nu=1}^{3n} \sum_{\vect{k}} \frac{1}{\sqrt{2Nm_j\omega_{\nu,\vect{k}}}} \,\bigl( \hat a_{\nu,\vect{k}} +\hat a^\dagger_{\nu,-\vect{k}} \bigr) \,e^{i\vect{k}\cdot\vect{x}_{lj}^0}\, \vect{\epsilon}_{\nu,\vect{k},j}\,,\label{eq:ulj}\\
\vect{p}_{lj} &=& i \sum_{\nu=1}^{3n} \sum_{\vect{k}} \sqrt{\frac{m_j\omega_{\nu,\vect{k}}}{2N}}\,\bigl(  \hat a^\dagger_{\nu,-\vect{k}} - \hat a_{\nu,\vect{k}} \bigr) \,e^{i\vect{k}\cdot\vect{x}_{lj}^0}\,\vect{\epsilon}_{\nu,\vect{k},j}\,,\label{eq:plj}
\eeq
where $\nu$ labels the phonon branch (of which there are $3n$ for a three-dimensional crystal with $n$ atoms per primitive cell), $\vect{k}$ labels the phonon momentum within the 1BZ, $N$ is the total number of primitive cells, $m_j$ is the mass of the $j$th atom in the primitive cell, and $\hat{a}^\dagger_{\nu, \vect{k}}, \hat{a}_{\nu, \vect{k}}$ are the phonon creation and annihilation operators. The phonon energies $\omega_{\nu,\vect{k}}$ and eigenvectors $\vect{\epsilon}_{\nu,\vect{k},j}=\vect{\epsilon}_{\nu,-\vect{k},j}^*$ are obtained by solving the eigensystem of $\tens{V}_{ll'jj'}$, for which we use the open-source code \texttt{phonopy}~\cite{Togo2015a}. The target Hamiltonian then reads
\begin{equation}
\hat H = \sum_{\nu=1}^{3n}\sum_{\vect{k}} \omega_{\nu,\vect{k}} \hat a^\dagger_{\nu,\vect{k}} \hat a_{\nu,\vect{k}} +\OO\left( \hat{a}^3 \right)\,.
\end{equation}

\begin{figure}[t]
\includegraphics[width=0.7\linewidth]{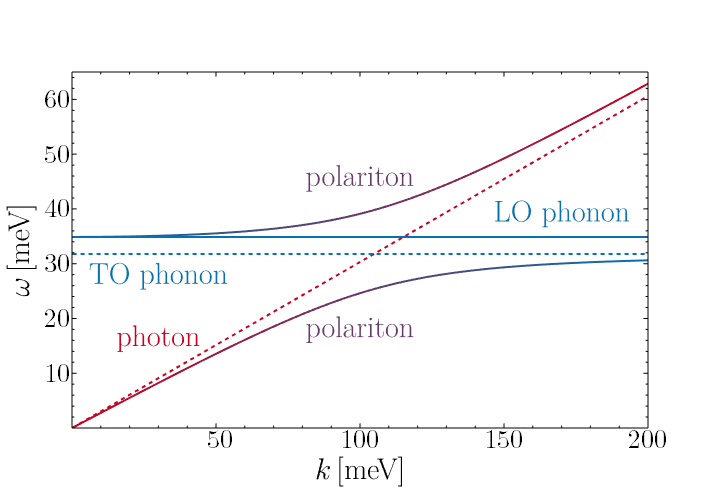}
\centering
\caption{Dispersion of phonon polaritons in GaAs near the center of the 1BZ, $k\sim\omega$. The mixing between the photon and TO phonons is maximal at $\omega\sim k$. At $k\ll\omega$, the TO phonon-like modes are degenerate with the LO phonon mode (blue line), while at $\omega\gg k$ they approach their unperturbed value (dotted blue line), and an LO-TO splitting is present.}
\label{fig:polariton}
\end{figure}

For a polar crystal, since the ions are electrically charged, some of the phonon modes mix with the photon. This mixing has a negligible impact in most of the 1BZ where $k\gg\omega$, and in particular does not affect the DM scattering calculations in Refs.~\cite{Knapen:2017ekk,Griffin:2018bjn,Trickle:2019nya,Griffin:2019mvc}. However, near the center of the 1BZ where $k\lesssim\omega$ -- relevant for DM absorption -- the photon-phonon mixing modifies the dispersions to avoid a level crossing. The true energy eigenstates are linear combinations of photon and phonon modes, known as phonon polaritons. This is shown in Fig.~\ref{fig:polariton} for gallium arsenide (GaAs) as a simple example. For an isotropic diatomic crystal like GaAs, the two degenerate, (mostly) transverse optical (TO) phonon modes at $k\gg\omega$ continue to photon-like modes at $k\ll\omega$, and vice versa. The phonon-like modes at $k\ll\omega$ do not have the same energies as away from the polariton regime: the TO phonon-like modes become degenerate with the longitudinal optical (LO) phonon mode at $\omega_\text{LO}$ as $k\to0$, whereas there is an LO-TO splitting, $\omega_\text{TO}\ne\omega_\text{LO}$, at $k\gg\omega$. For more complex crystals like sapphire (Al$_2$O$_3$), quartz (SiO$_2$) and calcium tungstate (CaWO$_4$), the mixing involves more phonon modes, and in general shift all their energies with respect to the eigenvalues $\omega_{\nu,\vect{k}}$ computed from diagonalizing just the lattice Hamiltonian.

To account for the photon-phonon mixing, we write the total Hamiltonian of electromagnetic fields coupling to the ions in the target crystal, and diagonalize its quadratic part via a Bogoliubov transformation. We explain this procedure in detail in Appendix~\ref{app:polariton}. The resulting diagonal Hamiltonian is
\begin{equation}\label{eq:diagHamPho}
\hat H = \sum_{\nu=1}^{3n+2}\sum_{\vect{k}} \omega'_{\nu,\vect{k}} \hat a'^\dagger_{\nu,\vect{k}} \hat a'_{\nu,\vect{k}} +\OO\left( \hat{a}'^3 \right) \,.
\end{equation}
At each $\vect{k}$, there are $(3n+2)$ modes, created (annihilated) by $\hat a'^\dagger_{\nu,\vect{k}}$ ($\hat a'_{\nu,\vect{k}}$), which are linear combinations of $3n$ phonon modes and 2 photon polarizations. Among them, 5 are gapless at $k=0$, including 3 acoustic phonons and 2 photon-like polaritons. The number of gapped modes, $3n-3$, is the same as in the phonon-only theory, but their energy spectrum is shifted, $\{\omega'_{\nu=(6,\dots,3n+2),\vect{k}}\} \ne \{\omega_{\nu=(4,\dots,3n),\vect{k}}\}$. The original phonon modes are linear combinations of the phonon polariton eigenmodes:
\begin{equation}
\hat a_{\nu,\vect{k}} = \sum_{\nu'=1}^{3n+2} \bigl( \mathbb{U}_{\nu\nu',\vect{k}}\, \hat a'_{\nu',\vect{k}} +\mathbb{V}_{\nu\nu',\vect{k}}\, \hat a^{\prime\dagger}_{\nu',-\vect{k}} \bigr)\,.
\label{eq:polmix}
\end{equation}

For DM coupling to the atomic displacements $\vect{u}_{lj}$, the perturbing potential is given by Eq.~\eqref{eq:fj} and therefore
\begin{eqnarray}
\hat{\delta H_0} | 0 \rangle &=& \sum_{lj} \sum_{\nu=1}^{3n} \sum_{\vect{k}} e^{i(\vect{p}-\vect{k})\cdot\vect{x}_{lj}^0} \frac{1}{\sqrt{2Nm_j\omega_{\nu,\vect{k}}}} \,\vect{f}_j \cdot \vect{\epsilon}_{\nu,\vect{k},j}^*\,\bigl( \hat a_{\nu,-\vect{k}} +\hat a^\dagger_{\nu,\vect{k}} \bigr) \,| 0 \rangle \nonumber\\
&=& \sqrt{\frac{N}{2}} \sum_{\nu=1}^{3n}\sum_{j} \frac{1}{\sqrt{m_j\omega_{\nu,\vect{p}}}} \,\vect{f}_j \cdot \vect{\epsilon}_{\nu,\vect{p},j}^*\,\bigl( \hat a_{\nu,-\vect{p}} +\hat a^\dagger_{\nu,\vect{p}} \bigr) \,| 0 \rangle \nonumber\\
&=& \sqrt{\frac{N}{2}} \sum_{\nu=1}^{3n}\sum_{\nu'=1}^{3n+2}\sum_{j} \frac{1}{\sqrt{m_j\omega_{\nu,\vect{p}}}}\,\vect{f}_j \cdot \vect{\epsilon}_{\nu,\vect{p},j}^* \,\bigl( \mathbb{U}_{\nu\nu',\vect{p}}^* + \mathbb{V}_{\nu\nu',-\vect{p}} \bigr) \,|\nu',\vect{p}\rangle\,,
\label{eq:dH00-pho}
\end{eqnarray}
where $|\nu',\vect{p}\rangle = \hat a'^\dagger_{\nu',\vect{p}} \,| 0 \rangle$. To arrive at the second equation, we have used the identity $\sum_l e^{i(\vect{p}-\vect{k})\cdot\vect{x}_{lj}^0}=N\delta_{\vect{k},\vect{p}}$ (for $\vect{k},\vect{p}\in\text{1BZ}$). It follows that
\begin{equation}
\langle \nu,\vect{k}\, | \hat{\delta H_0} | 0 \rangle = \delta_{\vect{k},\vect{p}}\sqrt{\frac{N}{2}} \sum_{\nu'=1}^{3n}\sum_{j} \frac{1}{\sqrt{m_j\omega_{\nu',\vect{p}}}}\,\vect{f}_j \cdot \vect{\epsilon}_{\nu',\vect{p},j}^* \,\bigl(\mathbb{U}_{\nu'\nu,\vect{p}}^* + \mathbb{V}_{\nu'\nu,-\vect{p}} \bigr)\,,
\end{equation}
where we have swapped the dummy indices $\nu$ and $\nu'$. The DM absorption rate per unit target mass, $R=\langle\Gamma\rangle/(N m_\text{cell})$, is therefore
\begin{equation}
R = \frac{2\omega}{m_\text{cell}}\int d^3v_a\, f(\vect{v}_a) \sum_{\nu=6}^{3n+2} \frac{\omega'_{\nu,\vect{p}}\gamma_{\nu,\vect{p}}}{(\omega^2-\omega_{\nu,\vect{p}}'^2)^2+(\omega\gamma_{\nu,\vect{p}})^2} \biggl|\sum_{j}\sum_{\nu'=1}^{3n} \frac{1}{\sqrt{m_j\omega_{\nu',\vect{p}}}}\, \vect{f}_j \cdot \vect{\epsilon}_{\nu',\vect{p},j}^* \bigl( \mathbb{U}_{\nu'\nu,\vect{p}}^* +\mathbb{V}_{\nu'\nu,-\vect{p}} \bigr) \biggr|^2\,,
\label{eq:Rpho}
\end{equation}
where $\omega = m_a$, $\vect{p} = m_a \vect{v}_a$, and $m_\text{cell}$ is the total mass of the atoms in a primitive cell. For our numerical calculations, we use the \texttt{phonopy} code~\cite{Togo2015a} to process DFT output~\cite{Griffin:2018bjn,Trickle:2019nya,Griffin:2019mvc} to obtain the unmixed phonon energies and eigenvectors $\omega_{\nu',\vect{p}}$, $\vect{\epsilon}_{\nu',\vect{p},j}$ as mentioned above, and then compute the polariton-corrected energy eigenvalues $\omega'_{\nu,\vect{p}}$ and mixing matrices $\mathbb{U}$, $\mathbb{V}$ via the algorithm of Refs.~\cite{Toth-Lake,Colpa1978a}. We relegate the technical details to Appendix~\ref{app:polariton}, and review the diagonalization algorithm~\cite{Toth-Lake,Colpa1978a} in Appendix~\ref{app:diag}.


\subsection{Magnon excitations}
\label{sec:magnon_rate_calc}

We now move to the case of magnons. The target Hamiltonian is a spin lattice model, with the following general form:
\begin{equation}
\hat H =\sum_{ll'jj'}\vect{S}_{lj} \cdot \tens{J}_{ll'jj'} \cdot \vect{S}_{l'j'}+\mu_B \vect{B} \cdot\sum_{lj}g_j\vect{S}_{lj}\,,
\label{eq:SpinHam}
\end{equation}
where $l,~l'$ label the magnetic unit cells and $j,~j'$ the magnetic ions inside the unit cell, $\mu_B = \frac{e}{2m_e}$ is the Bohr magneton, $\vect{B}$ is an external uniform magnetic field, and $g_j$ are the magnetic ions' Land\'e $g$-factors.
In the simplest case of the Heisenberg model, $\tens{J}_{ll'jj'}\propto\tens{1}$ for pairs of $lj$ and $l'j'$ on (nearest, next-to-nearest, etc.)\ neighboring sites. One material which is well described by this simple model is yttrium ion garnet (YIG)~\cite{Saga,Princep_2017}, which has already been considered for DM detection~\cite{Ruoso:2015ytk,Barbieri2017a,Crescini:2018qrz,Flower:2018qgb,Trickle:2019ovy,Chigusa2020a}. However, as we will see below, materials with spin-spin interactions beyond the simplest Heisenberg type can be useful for enhancing DM-magnon couplings.

The spin-spin interactions in Eq.~\eqref{eq:SpinHam} can result in a ground state with magnetic order. Here we focus on the simplest case of commensurate magnetic dipole orders, for which a rotation on each sublattice can take $\vect{S}_{lj}$ to a local coordinate system where each spin points in the $+\hat{\vect{z}}$ direction:
\begin{equation}\label{eq:SpinRot}
\vect{S}_{lj} = \tens{R}_j \cdot \vect{S}_{lj}' \,,
\qquad
\langle \vect{S}_{lj}'\rangle = ( 0,\, 0,\, S_j)\,.
\end{equation}
YIG and many other magnetic insulators have commensurate magnetic order. The calculation can be easily generalized to single-$Q$ incommensurate orders, as we discuss in Appendix~\ref{app:magnon}.

For a magnetically ordered system, the lowest energy excitations are magnons. To obtain the canonical magnon modes, we first apply the Holstein-Primakoff transformation to write the Hamiltonian in terms of bosonic creation and annihilation operators:
\begin{equation}\label{eq:HolPri}
S_{lj}'^+ = \bigl(2S_j-\hat a_{lj}^\dagger \hat a_{lj}\bigr)^{1/2}\, \hat a_{lj}\,,\qquad
S_{lj}'^- = \hat a_{lj}^\dagger\,\bigl(2S_j-\hat a_{lj}^\dagger \hat a_{lj}\bigr)^{1/2}\,,\qquad
S_{lj}'^z = S_j -\hat a_{lj}^\dagger \hat a_{lj}\,,
\end{equation}
where $S_{lj}'^\pm=S_{lj}'^x\pm iS_{lj}'^y$. The Holstein-Primakoff transformation ensures that the spin commutation relations $[S_{lj}'^\alpha, S_{l'j'}'^\beta] = \delta_{ll'}\delta_{jj'}\,i\epsilon^{\alpha\beta\gamma}S_{lj}'^\gamma$ are preserved when the usual canonical commutation relations $[\hat a_{lj}, \hat a_{l'j'}^\dagger]=\delta_{ll'}\delta_{jj'}$ are imposed. As in the phonon case, translation symmetry instructs us to go to momentum space:
\begin{eqnarray}
\hat a_{lj} &=& \frac{1}{\sqrt{N}} \sum_{\vect{k}} \hat a_{j,\vect{k}}\, e^{i\vect{k}\cdot\vect{x}_{lj}^0}\,,
\label{eq:Fourier}
\end{eqnarray}
where $\vect{k}\in\text{1BZ}$. The quadratic Hamiltonian, whose detailed form can be found in Appendix~\ref{app:magnon}, only couples modes with the same momentum, i.e.\ $\hat a_{j,\vect{k}}$ and $\hat a_{j',\vect{k}}$, $\hat a^\dagger_{j',-\vect{k}}$. A Bogoliubov transformation takes the quadratic Hamiltonian to the desired diagonal form:
\begin{equation}\label{eq:MagHamDia}
\hat{H} = \sum_{\nu=1}^n\sum_{\vect{k}} \omega_{\nu,\vect{k}} \hat a'^\dagger_{\nu,\vect{k}} \hat a'_{\nu,\vect{k}} +\OO(\hat a'^3)\,.
\end{equation}
At each $\vect{k}$, there are $n$ magnon modes with $n$ the number of spins per magnetic unit cell. These energy eigenmodes are created (annihilated) by $\hat a'^\dagger_{\nu,\vect{k}}$ ($\hat a'_{\nu,\vect{k}}$), which are related to the unprimed creation and annihilation operators by
\be
\hat a_{j,\vect{k}} = \sum_{\nu=1}^n \bigl( \mathbb{U}_{j\nu,\vect{k}}\, \hat a'_{\nu,\vect{k}} +\mathbb{V}_{j\nu,\vect{k}}\, \hat a^{\prime\dagger}_{\nu,-\vect{k}} \bigr)\,.
\ee

For DM coupling to the effective spins $\vect{S}_{lj}$, the interaction is given by Eq.~\eqref{eq:fj}, and we find, in complete analogy with Eq.~\eqref{eq:dH00-pho},
\begin{eqnarray}
\hat{\delta H_0} | 0 \rangle &=& \sum_{lj}\sum_{\vect{k}} e^{i(\vect{p}-\vect{k})\cdot\vect{x}_{lj}^0} \,\sqrt{\frac{S_j}{2N}}\, \vect{f}_j \cdot  (\vect{r}_j^*\hat a_{j,-\vect{k}} +\vect{r}_j\hat a_{j,\vect{k}}^\dagger)\, |0\rangle \nonumber\\
&=& \delta_{\vect{k},\vect{p}} \sqrt{\frac{N}{2}}\sum_{j} \sqrt{S_j}\, \vect{f}_j \cdot  (\vect{r}_j^*\hat a_{j,-\vect{k}} +\vect{r}_j\hat a_{j,\vect{k}}^\dagger)\, |0\rangle \nonumber\\
&=& \delta_{\vect{k},\vect{p}} \sqrt{\frac{N}{2}} \sum_{\nu=1}^n\sum_j \sqrt{S_j}\, \vect{f}_j \cdot \bigl( \mathbb{U}_{j\nu,\vect{p}}^*\vect{r}_j +\mathbb{V}_{j\nu,-\vect{p}} \vect{r}_j^* \bigr)\, |\nu,\vect{p}\rangle\,,
\end{eqnarray}
where $\vect{r}_j \equiv (\tens{R}_j^{xx},\, \tens{R}_j^{yx},\, \tens{R}_j^{zx})+i\, (\tens{R}_j^{xy},\, \tens{R}_j^{yy},\, \tens{R}_j^{zy})$, and $|\nu,\vect{p}\rangle = \hat a'^\dagger_{\nu,\vect{p}} \,| 0 \rangle$. Therefore,
\begin{equation}
\langle \nu, \vect{k} | \hat{\delta H_0} | 0 \rangle = \delta_{\vect{k},\vect{p}} \sqrt{\frac{N}{2}} \sum_{\nu=1}^n\sum_j \sqrt{S_j}\, \vect{f}_j \cdot \bigl( \mathbb{U}^*_{j\nu,\vect{p}}\vect{r}_j +\mathbb{V}_{j\nu,-\vect{p}} \vect{r}_j^* \bigr)\,.
\end{equation}
We can now obtain the DM absorption rate per unit target mass:
\begin{equation}
R = \frac{2\omega}{m_\text{cell}}\int d^3v_a\, f(\vect{v}_a) \sum_{\nu=n_0+1}^{n}\frac{\omega_{\nu,\vect{p}}\gamma_{\nu,\vect{p}}}{(\omega^2-\omega_{\nu,\vect{p}}^2)^2+(\omega\gamma_{\nu,\vect{p}})^2}\, \biggl|\sum_j \sqrt{S_j}\, \vect{f}_j \cdot\bigl( \mathbb{U}^*_{j\nu,\vect{p}}\vect{r}_j +\mathbb{V}_{j\nu,-\vect{p}} \vect{r}_j^* \bigr) \biggr|^2 \,,
\label{eq:Rmag}
\end{equation}
where $n_0$ is the number of gapless modes, which depends on the material (in particular, on the symmetry breaking pattern). Similarity to the phonon formula Eq.~\eqref{eq:Rpho} is apparent. We again use the algorithm of Refs.~\cite{Toth-Lake,Colpa1978a}, reviewed in Appendix~\ref{app:diag}, to solve the diagonalization problem to obtain the magnon energies $\omega_{\nu,\vect{p}}$ and mixing matrices $\mathbb{U}$, $\mathbb{V}$.


\section{Selection rules and ways around them}
\label{sec:selection}

Depending on the DM couplings $\vect{f}_j$, excitation rates for some of the phonon or magnon modes can be suppressed. In the context of DM scattering, it has been known that acoustic and optical phonons are sensitive to different types of DM couplings~\cite{Trickle:2019nya,Griffin:2019mvc}: if DM couples to the inequivalent atoms/ions with the same sign (different signs), the single phonon excitation rate is dominated by acoustic (optical) phonons, corresponding to in-phase (out-of-phase) oscillations of the atoms/ions. 

The same considerations apply to absorption of DM, though here the gapless acoustic phonons are kinematically inaccessible, and therefore only gapped optical phonons can be excited. Thus, the rate is suppressed if all $\vect{f}_j$ point in the same direction. As an extreme example, consider $\vect{f}_j = m_j \vect{f}$, with $\vect{f}$ a constant vector. Up to the photon-phonon mixing (which mostly shifts the energy eigenvalues while leaving the factor $( \mathbb{U}_{\nu'\nu,\vect{p}}^* +\mathbb{V}_{\nu'\nu,-\vect{p}} )$ close to $\delta_{\nu'\nu}$), the rate in Eq.~\eqref{eq:Rpho} is proportional to $\bigl|\sum_j \sqrt{m_j} \vect{f}\cdot\vect{\epsilon}_{\nu,\vect{p},j}^*\bigr|^2$. However, one can show from translation symmetry that $\sqrt{m_j}\vect{f}$ can be written as a linear combination of the polarization vectors $\vect{\epsilon}_{\nu,\vect{p}\to\vect{0},j}$ with $\nu\in$ acoustic --- see for example the explicit discussion of~\cite{Cox2019a}, consistent with the earlier results in Refs.~\cite{Knapen:2017ekk,Griffin:2018bjn}. This is not surprising since gapless acoustic phonons are Goldstone modes of the broken translation symmetries. Thus, by the orthogonality of the phonon polarization vectors, optical phonons do not contribute to the rate in the $\vect{p}\to\vect{0}$ limit, and only higher order terms in the DM velocity can give a nonzero contribution. In the next section, in the context of axion DM, we will encounter both cases where the suppression due to the $\vect{f}_j$'s being aligned is present and absent, and will identify a process free of the suppression as a viable detection channel.

Additional selection rules may be present among the optical phonons. For example, sapphire has 27 optical phonon branches, but we find that near the 1BZ center, only 10 of them couple to the axion-induced electric field (in the presence of an external magnetic field). Furthermore, 8 of the 10 modes are degenerate in pairs, reducing the total number of distinct resonances to 6, as seen in Fig.~\ref{fig:spectrum}. This is consistent with the well-known fact that, due to crystalline symmetries, sapphire has 6 infrared-active phonon modes~\cite{Bhagavantam1939,PhysRev.132.1474,PhysRevB.61.8187}.  Thus, despite the existence of many optical phonon modes, sapphire does not really offer broadband coverage of the axion mass. The same observation, that only a subset of gapped phonon modes couple to axion DM, also holds for the other targets considered: \ce{SiO2} and \ce{CaWO4} -- see Fig.~\ref{fig:spectrum} (GaAs has only 3 optical phonon modes which are all degenerate and can couple to axion DM). To broaden the mass coverage, it is therefore necessary to run experiments with several target materials with distinct phonon frequencies.

There are also selection rules in the case of magnon excitations. It has been pointed out that, assuming the absence of an external magnetic field, for a target system described by the Heisenberg model with quenched orbital angular momentum, such as YIG, only gapless magnons can be excited in the zero momentum transfer limit~\cite{Trickle:2019ovy,Tabuchi2015a}. To understand why, let us review and quantify the semiclassical argument given in Ref.~\cite{Trickle:2019ovy}. Within a coherence length, the DM field couples to the spins as a uniform magnetic field, causing all the spins to precess. As a result, the rate of change in the Heisenberg interaction energy between any pair of spins is proportional to
\begin{eqnarray}
\frac{d}{dt} \bigl(\vect{S}_{lj}\cdot\vect{S}_{l'j'}\bigr) &=& \frac{d\vect{S}_{lj}}{dt}\cdot\vect{S}_{l'j'} + \vect{S}_{lj}\cdot \frac{d\vect{S}_{l'j'}}{dt} = (\vect{f}_j\times\vect{S}_{lj})\cdot\vect{S}_{l'j'} + \vect{S}_{lj}\cdot(\vect{f}_{j'}\times\vect{S}_{l'j'}) \nonumber\\
&=& (\vect{f}_j-\vect{f}_{j'})\cdot (\vect{S}_{lj}\times\vect{S}_{l'j'})\,,
\end{eqnarray}
which vanishes for $\vect{f}_j=\vect{f}_{j'}$. Therefore, if the target system is described by the Heisenberg Hamiltonian, and all $\vect{f}_j$ are equal (which is quite generic since they all originate from DM-electron spin coupling), the total energy cannot change in response to the DM field, and no gapped magnons can be excited. In other words, the DM field only couples to gapless magnons. In the case of scattering, the rate is not severely suppressed by this fact since the scattering kinematics allows access to finite momentum magnons, and hence sufficient energy deposition to be detected.

The situation is much worse in the case of absorption, because the momentum transfer is small in comparison to the DM mass due to its small velocity $v \sim 10^{-3}$. Therefore, gapless modes cannot be excited due to kinematics. As a result, the detection rate is severely suppressed by powers of DM velocity. We have checked this explicitly for several target materials. In what follows, let us expand on the two examples.

\paragraph{YIG (\ce{Y3Fe5O12}).} The crystal primitive cell of YIG consists of four copies of \ce{Y3Fe5O12}. The magnetic ions are Fe$^{3+}$, each of which has spin 5/2. The magnetic unit cell coincides with the crystal primitive cell, and contains 20 magnetic ions. The spin Hamiltonian has antiferromagnetic Heisenberg interactions, which we include up to third nearest neighbors~\cite{Princep_2017}. The ground state has ferrimagnetic order, where the 12 magnetic ions on the tetrahedral sites and the 8 magnetic ions on the octahedral sites have spins pointing in opposite directions~\cite{Saga}, taken to be $\pm \hat z$. The symmetry breaking pattern is SO(3)$\,\to\,$SO(2), and hence there are two broken generators $S_x$, $S_y$. There is however just one Goldstone mode (with quadratic dispersion) due to the nonvanishing expectation value of the commutator between the broken generators, $\langle[S_x, S_y]\rangle=(i/2)\langle S_z\rangle\ne 0$~\cite{Nielsen:1975hm,Watanabe:2011ec,Watanabe:2012hr,Watanabe:2014fva}. Thus, among the 20 magnon branches, only one is gapless. We find that at zero momentum, the $j$ sum in the rate formula Eq.~\eqref{eq:Rmag}, $\sum_j\sqrt{S_j}\cdot\bigl(\mathbb{U}^*_{j\nu,\vect{0}}\vect{r}_j +\mathbb{V}_{j\nu,\vect{0}} \vect{r}_j^*\bigr)$, indeed vanishes for all but the gapless mode ($\nu=1$), confirming the argument above that the DM field only couples to gapless magnons.

\paragraph{\ce{Ba3NbFe3Si2O14}.} This is an example of materials with incommensurate magnetic order. We discuss the generalization needed in the rate calculation in Appendix~\ref{app:magnon}, with the final result given in Eq.~\eqref{eq:Rmag-i}. The magnetic unit cell contains three magnetic ions Fe$^{3+}$ with spin 5/2, which form a triangle in the $x$-$y$ plane. The crystal consists of layers stacked in the $z$ direction. The antiferromagnetic Heisenberg interactions result in a frustrated order with 120$^\circ$ between the three spins that are nearest neighbors. Further, the chiral structure of inter-layer Heisenberg exchange couplings results in a rotation of the order in the $z$ direction, with a wavevector that is irrational, $\vect{Q} \simeq 0.1429 \,(2\pi/c)\,\hat{z}$ where $c\simeq 5.32\,$\AA\ is the inter-layer lattice spacing~\cite{Toth-Lake}. This is known as a single-$Q$ incommensurate order. All 3 generators of SO(3) are broken while the ground state has zero total magnetization, so there are 3 Goldstone modes. These appear at $\vect{k}=\vect{0},\pm\vect{Q}$, which are also the momenta near which the axion coupling is nonzero due to (generalized) momentum conservation. We find that at all three momenta, the $j$ sum in the generalized rate formula Eq.~\eqref{eq:Rmag-i} is nonzero only for $\nu=1$, {\em i.e.}\ the gapless modes, again confirming the argument above that the DM field only couples to gapless magnons. 

Nevertheless, there are several possibilities to alleviate the problem. First, one can consider targets involving additional, non-Heisenberg interactions. These additional terms can explicitly break the rotational symmetries, causing the otherwise gapless Goldstone modes to become gapped, and match the DM absorption kinematics. Concretely, we can identify two ways of implementing this idea:
\begin{itemize}
\item An {\it external magnetic field} $\vect{B}\ne0$ can generate a gap for the lowest magnon branch equal to the Larmor frequency,
\begin{equation}
\omega_L = 2\mu_B B = 0.12\,\text{meV}\,\biggl(\frac{B}{\text{T}}\biggr)\,,
\label{eq:omegaL}
\end{equation}
assuming $g_j=2$ for all $j$. The QUAX experiment~\cite{Ruoso:2015ytk,Barbieri2017a,Crescini:2018qrz} makes use of this to search for axion DM with $m_a\sim\OO(0.1\,\text{meV})$, currently in the regime where the magnon number is large and a classical description can be used. Recently, a calculation to exploit this effect in the quantum regime, similar to the derivation in Sec.~\ref{sec:magnon_rate_calc}, was carried out in Ref.~\cite{Chigusa2020a}. However, sensitivity of such a setup is limited to sub-meV DM by the achievable magnetic field strengths, and only the lowest magnon mode(s) can be excited.
\item There are materials with {\it anisotropic interactions} where the number of gapless Goldstone modes is reduced. In Sec.~\ref{sec:sensitivity-mag}, we consider a concrete example, \ce{NiPS3}, where the $\tens{J}_{ll'jj'}$ matrices in the spin Hamiltonian Eq.~\eqref{eq:SpinHam} have unequal diagonal entries~\cite{Kim_2019}. In this case, two gapped magnon modes at 12\,meV and 44\,meV can couple to axions. There are also materials with nonzero off-diagonal entries in $\tens{J}_{ll'jj'}$, arising from {\em e.g.}\ Dzyaloshinskii-Moriya interactions, that could be used to achieve the same result.
\end{itemize}
An orthogonal route to solve the problem is to use targets where the DM-spin couplings $\vect{f}_j$ are nondegenerate.
\begin{itemize}
\item For materials with {\it nondegenerate Land\'e $g$-factors}, even a uniform magnetic field can drive the magnetic ions differently and excite gapped magnon modes; the same is true for a uniform DM field. The basic reason for this is the presence of spin-orbit couplings that break the degeneracy between the DM couplings to the magnetic ions' total effective spins.  Concretely, the Land\'e $g$-factors are given by
\begin{equation}
g_j=\frac{3}{2}+\frac{1}{2} \frac{s_j(s_j+1)-\ell_j(\ell_j+1)}{S_j(S_j+1)}\,,
\end{equation}
where $s_j$ and $\ell_j$ are respectively the spin and orbital angular momentum components of the total effective spin $\vect{S}_{lj}$. In the simplest and most common case of magnetic ions with quenched orbital angular momenta (\emph{i.e.} $\ell_j\simeq 0$), we recover the usual result $g_j=2$. Breaking the degeneracy requires the magnetic ions to have different spin and orbital angular momentum compositions. We demonstrate how this allows for axion couplings to gapped magnons in Sec.~\ref{sec:sensitivity-mag}.
\end{itemize}

As in the case of phonons, there are usually additional selection rules due to crystalline symmetries. As a result, the strategies discussed above usually generate axion couplings to only a subset of gapped magnon modes -- see Fig.~\ref{fig:spectrum}. Therefore, multiple target materials which cover complementary ranges of magnon frequencies are desirable.


\section{Axion couplings and detection channels}
\label{sec:couplings}

The derivation and discussion in the previous two sections apply to general field-like DM candidates. We now specialize to the case of axion DM. Our goal in the present section is to identify the most promising detection channels involving phonon or magnon excitation via order-of-magnitude estimates. We then examine these processes quantitatively in the next section.

The axion couplings of interest are already given in Eqs.~\eqref{eq:La} and \eqref{eq:non_rel_int_ham}. For the QCD axion, we have~\cite{Tanabashi:2018oca}
\begin{eqnarray}
g_{a\gamma\gamma} &=& C_\gamma \frac{\alpha}{2\pi f_a} = 2.03\times 10^{-12}\, C_\gamma \biggl(\frac{m_a}{10\,\text{meV}}\biggr) \GeV^{-1} \,,\label{eq:gayy} \\
g_{aff} &=& C_f \frac{m_f}{f_a} = 1.18\times 10^{-13}\, C_f \left(\frac{m_f}{\MeV}\right)\left(\frac{m_a}{\meV}\right) \,, \label{eq:gaff}\\
g_{an\gamma} &=& -g_{ap\gamma} = (3.7\pm 1.5)\times 10^{-3} \frac{1}{f_a} \frac{1}{\text{GeV}} = (6.5 \pm 2.6) \times 10^{-12} \biggl(\frac{m_a}{10\,\text{meV}}\biggr) \,\text{GeV}^{-2} \,,
\end{eqnarray}
where we have denoted $C_\gamma \equiv E/N - 1.92(4)$ with $E/N= 0$ ($8/3$) in the KSVZ (DFSZ) model. The axion-fermion couplings $C_f$ are also model dependent. In particular, the axion-electron coupling is $C_e=\sin^2\beta/3$ in the DFSZ model, where $\tan\beta$ is the ratio of the vacuum expectation values of the two Higgs doublets giving masses to the up and down-type quarks. In the KSVZ model, on the other hand, $C_e$ is $\OO(\alpha^2)$ suppressed. 

In the following subsections we consider axion couplings independent of external fields and in the presence of a magnetic field.\footnote{An external electric field shifts the equilibrium positions of the ions such that there is no net electric field at the new equilibrium positions, so it does not generate new axion couplings at leading order.} In each case, we discuss the phonon and magnon excitation processes that are allowed, and identify those with potentially detectable rates. The results of this exercise are summarized in Table~\ref{tab:chan}.


\subsection{Axion couplings independent of external fields}

The axion wind coupling to electron spin leads to a coupling to the spin component of $\vect{S}_{lj}$. From $\vect{s}_{lj}+\vect{\ell}_{lj} = \vect{S}_{lj}$ and $2\vect{s}_{lj}+\vect{\ell}_{lj} = g_j\vect{S}_{lj}$, we see that the axion wind couples to $\vect{s}_{lj} = (g_j-1)\, \vect{S}_{lj}$. Thus,
\begin{equation}
\delta \hat{ H } = -\frac{g_{aee}}{m_e} \nabla a\cdot\sum_{lj}(g_j-1)\vect{S}_{lj} = -\frac{g_{aee}}{m_e} (i\, m_a \vect{v}_a) \frac{a_0}{2} \cdot \sum_{lj}(g_j-1)\,\vect{S}_{lj}\, e^{i\vect{p}\cdot\vect{x}_{lj}^0-i\omega t} +\text{h.c.}
\end{equation}
In the notation of Eq~\eqref{eq:fj}, we thus have
\begin{equation}\label{eq:magWcoupling}
\vect{f}_j = - \frac{i}{\sqrt{2}} \,g_{aee} \,(g_j-1) \,\frac{\sqrt{\rho_a}}{m_e} \, \vect{v}_a \,.
\end{equation}
For an order of magnitude estimate of the rate, let us note that the mixing matrices $\mathbb{U}$, $\mathbb{V}$ in Eq.~\eqref{eq:Rmag} generically scale as $1/\sqrt{n}$ with $n$ the number of magnetic ions in a primitive cell. The maximum rate is obtained on resonance, which is parametrically given by
\begin{equation}
R \sim  \frac{g_{aee}^2 \,\rho_a v_a^2}{m_e^2}\frac{n_s}{\rho_T \gamma}
\sim (\text{kg$\cdot$yr})^{-1} \, \biggl(\frac{g_{aee}}{10^{-15}}\biggr)^2 \biggl(\frac{\mu\text{eV}}{\gamma}\biggr)\,.
\end{equation}
where $n_s$ and $\rho_T$ are the spin and mass densities of the target, taken to be $(5\,\text{\AA})^{-3}$ and 5\,g/cm$^3$, respectively (close to the values for YIG), in the estimate. We see that, with single magnon sensitivity, interesting values of $g_{aee}$ may be reached with less than a kilogram-year exposure.

The axion wind also couples to nucleon spins. However, these couplings do not excite magnons, since magnetic order originates from electron-electron interactions, meaning that the effective spins of magnetic ions that appear in the spin Hamiltonian Eq.~\eqref{eq:SpinHam} come from electrons. On the other hand, if the nuclear spins $\vect{S}_N$ are ordered ({\it e.g.}\ by applying an external magnetic field which does not affect the axion-nucleon couplings) and form a periodic structure, the axion wind couplings could excite phonons. However, the rate suffers from multiple suppressions. First, coupling to atomic displacements relies on the spatial variation of $\nabla a\cdot \vect{S}_N$, which brings in an additional factor of $v_a$ on top of the gradient: $\vect{f}_j \sim (C_f/f_a)\, m_a^2 a \,(\vect{v}_a\cdot\vect{S}_{N,j})\,\vect{v}_a$. Second, there is a further suppression for exciting optical phonons since $\vect{f}_j$ are approximately aligned with acoustic phonon polarizations if all $\vect{S}_{N,j}$ point in the same direction (see discussion in Sec.~\ref{sec:selection}). Even without taking into account the second suppression, the estimated on-resonance rate using Eq.~\eqref{eq:Rpho},
\begin{align}
R\sim\frac{C_f^2\rho_a m_a v_a^4}{f_a^2} \frac{n_s^2}{\rho_T^2\gamma_\nu} \sim (\text{kg$\cdot$yr})^{-1}\, C_f^2 \,\left(\frac{100\,\text{GeV}}{f_a}\right)^2 \left(\frac{m_a}{10\,\text{meV}}\right) \left(\frac{\mu\text{eV}}{\gamma_\nu}\right) \, ,
\end{align}
can be sizable only for uninterestingly low $f_a$. Therefore, we conclude that axion wind couplings to nucleon spins do not offer a viable detection channel.

Beyond the axion wind couplings, the axion field also turns magnetic dipole moments from $\vect{s}_f$ into oscillating EDMs, and one may consider phonon and magnon excitation by the resulting electromagnetic radiation fields. In the case of the electron, since the EDM coupling is perturbatively generated by the $aF\tilde F$ coupling, the process mentioned above is essentially converting the crystal magnetic field into electromagnetic radiation, and should be less efficient than applying an external magnetic field (discussed below in Sec.~\ref{sec:couplings-b}). In the case of nucleons, the EDM coupling in Eq.~\eqref{eq:La} is not much larger than the perturbative contribution from $aF\tilde F$, so the same conclusion applies.

In sum, for axion couplings independent of external fields, we have identified {\it magnon excitation via the axion wind coupling to electrons} as the only viable detection channel.


\subsection{Axion couplings in a magnetic field}
\label{sec:couplings-b}

In the presence of a DC magnetic field $\vect{B}$, the axion field induces oscillating electromagnetic fields via the $aF\tilde F$ coupling. Solving the modified Maxwell equations (see \textit{e.g.}\ Ref.~\cite{Visinelli2013a,Millar:2016cjp}), we find the induced electric field is $\vect{E}_a = -g_{a\gamma\gamma} a \,  \vect{\varepsilon}_\infty^{-1} \cdot \vect{B}$. Note that the high frequency dielectric constant, $\vect{\varepsilon}_\infty$, which takes into account screening effects from the fast-responding electrons while excluding ionic contributions, should be used when solving the macroscopic Maxwell equations --- essentially, since we are concerned with how the axion induced electric field acts on the ions, the ion charges should appear in the source term rather than being coarse grained into a macroscopic electric polarization. In the long-wavelength limit, the axion-induced electric field $\vect{E}_a$ couples to charged ions via an effective dipole coupling:
\begin{equation}
\delta \hat{H} = -e \sum_{lj} \vect{E}_a \cdot \tens{Z}_{j}^* \cdot \vect{u}_{lj} = e g_{a\gamma\gamma}a\, \sum_{lj} \vect{B}_a \cdot \vect{\varepsilon}_\infty^{-1} \cdot \tens{Z}_{j}^* \cdot \vect{u}_{lj} \,,
\end{equation}
where $\tens{Z}_{j}^*$ is the Born effective charge tensor of the $j$th ion in the primitive cell --- it captures the change in macroscopic polarization due to a lattice displacement, $\delta\vect{P} = e\,\tens{Z}_j^* \cdot \delta\vect{u}_{lj}/ \Omega$, and is numerically close to the ionic charge, $\tens{Z}_{j}^*\simeq Q_j \mathbb{1}$. It follows that
\begin{equation}\label{eq:phoBcoupling}
\vect{f}_j = \frac{1}{\sqrt{2}}\,g_{a\gamma\gamma}\frac{e \sqrt{\rho_a}}{m_a}\, \vect{B} \cdot \vect{\varepsilon}^{-1}_\infty \cdot \tens{Z}_j^* \,.
\end{equation}
Noting that the phonon polarization vectors scale as $1/\sqrt{n}$ with $n$ the number of ions in the primitive cell, and assuming photon-phonon mixing gives just a small correction, we can estimate the on-resonance rate from Eq.~\eqref{eq:Rpho} as follows:
\begin{equation}\label{eq:Rphoest}
R \sim \frac{g_{a\gamma\gamma}^2\rho_a}{m_a^3} \frac{Z^{*2}e^2B^2}{\varepsilon_\infty^2 m_\text{ion}^2\gamma} \sim (\text{kg$\cdot$yr})^{-1} \,\biggl(\frac{g_{a\gamma\gamma}}{10^{-13}\,\text{GeV}^{-1}}\biggr)^2 \biggl(\frac{100\,\text{meV}}{m_a}\biggr)^3 \biggl(\frac{B}{10\,\text{T}}\biggr)^2\biggl(\frac{\text{meV}}{\gamma}\biggr)\,,
\end{equation}
where we have taken $Z^*/\varepsilon_\infty\sim 1$ and $m_\text{ion}\sim 20\,$GeV. We see that, with single phonon sensitivity, there is excellent potential for reaching the QCD axion coupling if the axion mass is close to a phonon resonance. 

The axion-induced magnetic field is much smaller, $B_a\sim E_a v_a$. An order of magnitude estimate tells us that the magnon excitation rate by $B_a$ is much smaller than the phonon excitation rate by $E_a$: $\frac{R_\text{magnon}}{R_\text{phonon} }\sim \bigl(\frac{\mu_B B_a S_{lj}}{eZ^* E_a u_{lj}}\bigr)^2 \sim \frac{m_j m_a v_a^2}{m_e^2} \sim 10^{-8}$ for a 100\,meV axion.

In sum, we have identified {\it phonon excitation via the axion-photon coupling in a magnetic field} as the only novel viable detection channel when considering an external DC magnetic field.


\section{Projected sensitivity}
\label{sec:sensitivity}

We now compute the projected sensitivity for the two detection channels identified in the previous section (see Table~\ref{tab:chan}). In both phonon and magnon calculations, an important but elusive parameter is the resonance width, $\gamma_{\nu,\vect{p}}$, of each mode. While all other material parameters entering the rate calculation (equilibrium positions, phonon energies and eigenvectors, magnon energies and mixing matrices) can be computed within the quadratic Hamiltonian, the widths involve anharmonic interactions and are not always readily available. In what follows, we present the reach with reasonable assumptions for $\gamma_{\nu,\vect{p}}$. Our goal here is to demonstrate the viability of phonons and magnons for detecting axion DM, and motivate further study in the condensed matter and materials science community on phonon and magnon interactions. This will be crucial both for having more accurate inputs to the DM detection rate calculation, and for designing detectors to read out these excitations.


\subsection{Phonon excitation via the axion-photon coupling}

When axion absorption excites a phonon polariton, it cascade decays to a collection of lower energy phonons and photons on a timescale of $\gamma^{-1} \sim \text{ps} \,(\text{meV}/\gamma)$. Since the polariton is a phonon-like state which decays via anharmonic phonon couplings, theoretically it is most efficient to read out phonons (heat) in the final state. However, phonon readout (\textit{e.g.}\ through a transistor edge sensor) is complicated by the strong external magnetic field needed for the axion absorption process.  One possibility is to detect the phonon via evaporation of helium atoms deposited on the surface of the crystal.  The helium atoms are then detected well away from the crystal, such that the magnetic field is isolated from the sensor.\footnote{We thank Stephen Lyon and Thomas Schenkel for discussions on this experimental avenue.}  Alternatively, if a target material can be found in which the photon yield from phonon polariton decays is substantial, photon readout becomes a viable option. In this case, the photons may be focused by a mirror and lens and directed by a waveguide onto a single photon detector (\textit{e.g.}\ a superconducting nanowire) placed in a region of zero magnetic field. Both the phonon and photon readout possibilities sketched above will be studied in future work. In what follows, we simply assume 3 single phonon polariton excitation events per kilogram-year exposure (corresponding to 90\% C.L.\ for a background-free counting experiment) when presenting the reach.

\begin{figure}[t]
\centering
	\includegraphics[width=0.95\textwidth]{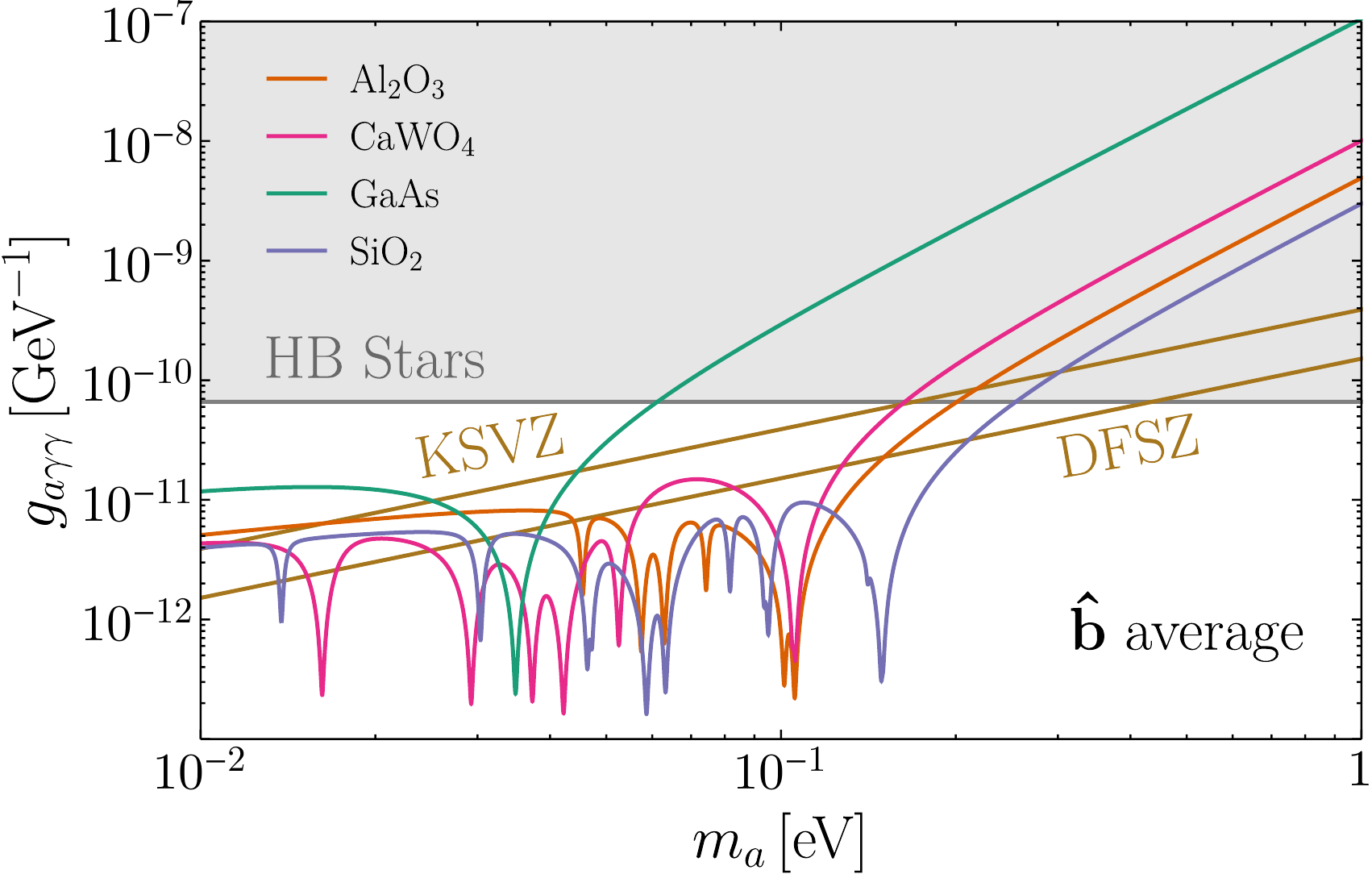}
	\caption{	\label{fig:phoreach1}
Projected reach on $g_{a \gamma \gamma}$ from axion absorption onto phonon polaritons in \ce{Al2O3}, \ce{CaWO4}, \ce{GaAs} and \ce{SiO2}, in an external 10\,T magnetic field, averaged over the magnetic field directions, assuming 3 events per kilogram-year. Also shown are predictions of the KSVZ and DFSZ QCD axion models, and horizontal branch (HB) star cooling constraints~\cite{Ayala2014a}.}
\end{figure}

\begin{figure}[t]
	\includegraphics[width=0.45\textwidth]{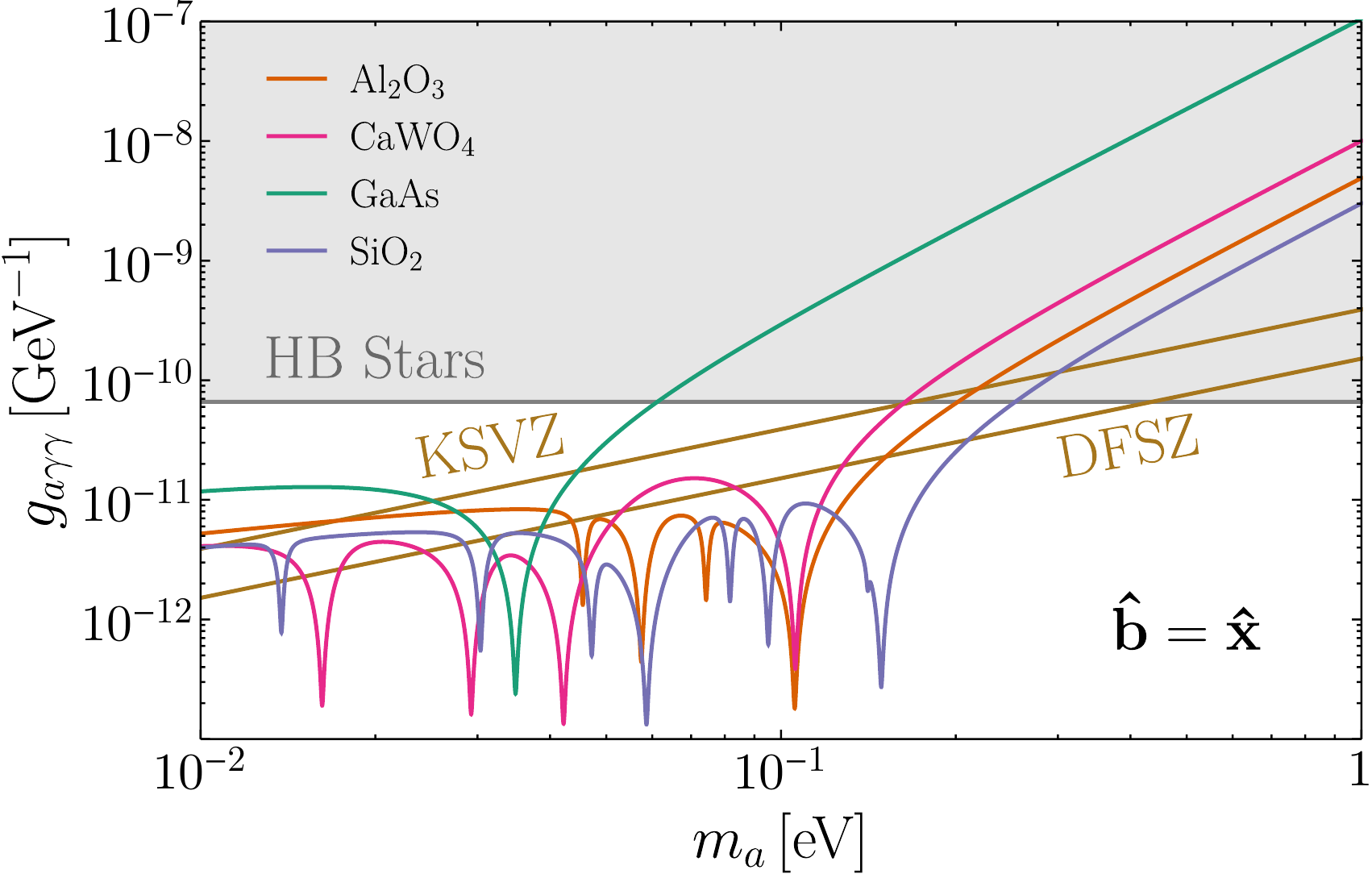}\qquad
	\includegraphics[width=0.45\textwidth]{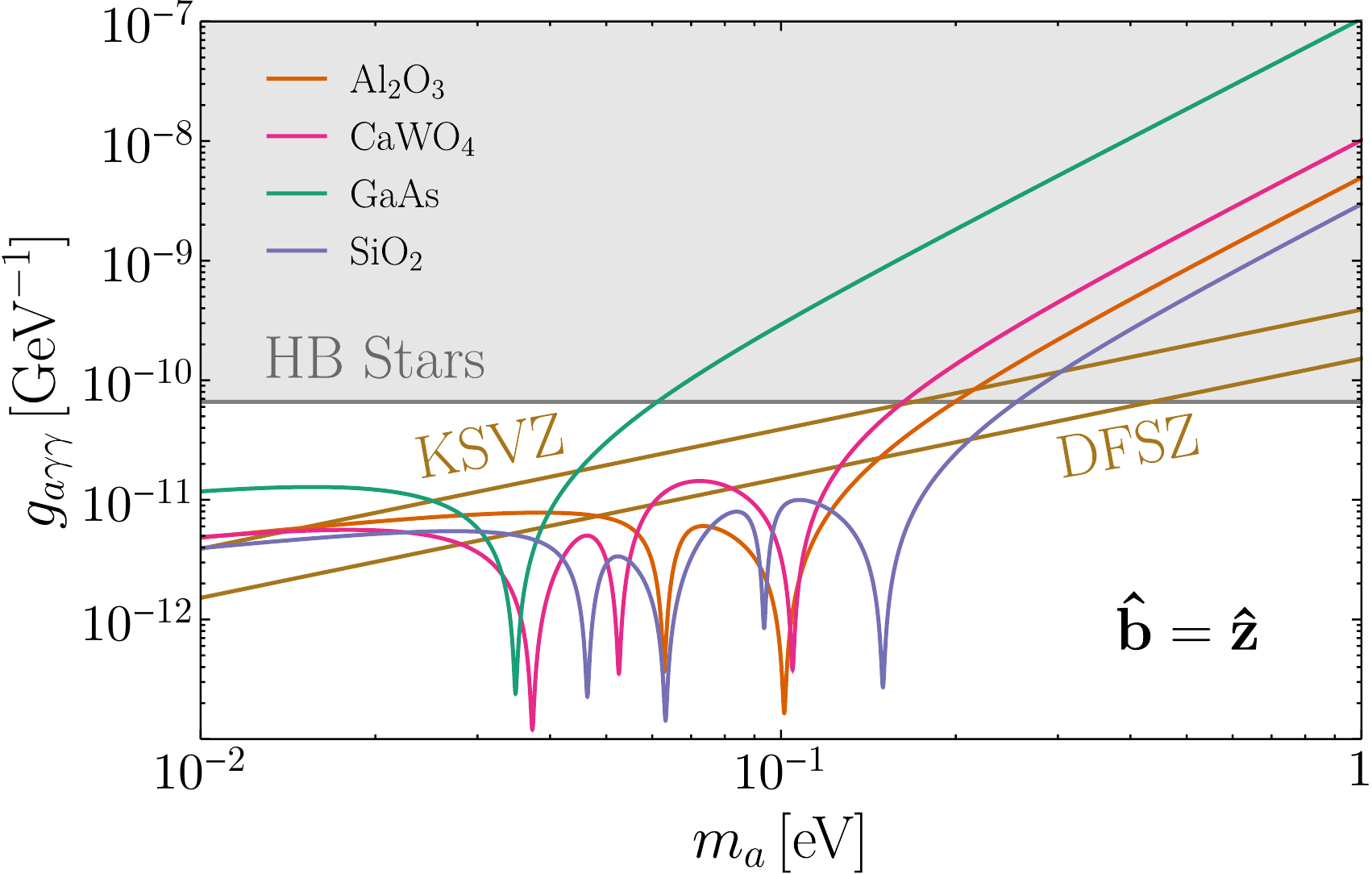}
	\caption{Similar to Fig.~\ref{fig:phoreach1}, but with the external magnetic field oriented in the $\vect{\hat{x}} \,  (\vect{\hat{z}})$ direction in the left (right) panel. The strength of axion-phonon couplings depends on the orientation of the magnetic field, and different resonances can be selected by changing the magnetic field direction.}
	\label{fig:phoreach2}
\end{figure}

To compute the projected sensitivity to the axion-photon coupling, we use the rate formula \Eq{eq:Rpho}, with the effective couplings $\vect{f}_j$ given in \Eq{eq:phoBcoupling}. We consider several example target materials and different orientations of the external magnetic field. The results are shown in Figs.~\ref{fig:phoreach1} and \ref{fig:phoreach2}. In these plots, we take the resonance widths $\gamma/\omega = 10^{-2}$, consistent with the order of magnitude of the measured numbers in sapphire. As we discussed in Sec.~\ref{sec:selection}, materials with different phonon frequencies play important complementary roles in covering a broader axion mass range.

It is also worth noting that, since the effective couplings $\vect{f}_j$ depend on the direction of the magnetic field $\mathbf{\hat{b}}$, the strengths of the resonances vary as $\mathbf{\hat{b}}$ is changed, as we can see in Fig.~\ref{fig:phoreach2}. For example, for a sapphire target, when $\mathbf{\hat{b}}$ is parallel (perpendicular) to the crystal $c$-axis, chosen to coincide with the $z$-axis here, only 2 (4) out of the 6 resonances appear. This observation provides a useful handle to confirm a discovery by running the same experiment with the magnetic field applied in different directions.



\subsection{Magnon excitation via the axion wind coupling}
\label{sec:sensitivity-mag}

\begin{figure}[t]
\centering
\includegraphics[width=0.95\textwidth]{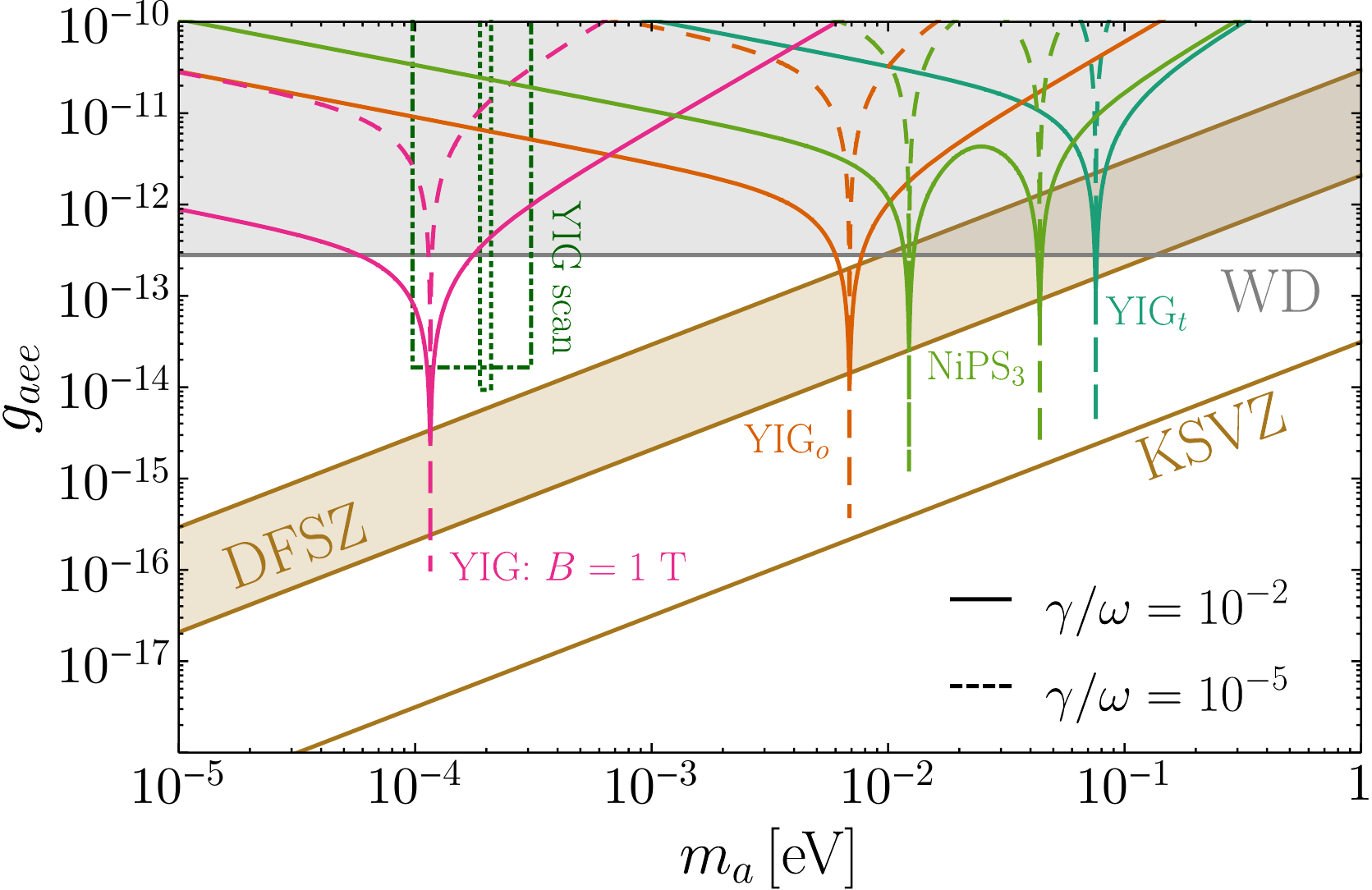}
\caption{Projected reach on $g_{aee}$ from axion-to-magnon conversion, compared with DFSZ (assuming $0.28 \le \tan{\beta} \le 140$) and KSVZ model predictions, as well as white dwarf (WD) constraints from Ref.~\cite{MillerBertolami2014a}. The suppression of axion-magnon couplings is alleviated by using the three strategies discussed in the main text: lifting gapless magnon modes by an external magnetic field (YIG target in a 1\,T magnetic field, compared to the scanning scheme of Ref.~\cite{Chigusa2020a}), anisotropic interactions (NiPS$_3$ target), and using targets with nondegenerate $g$-factors (hypothetical toy models based on YIG, referred to as YIG$_o$ and YIG$_t$). For all the cases considered we assume 3 events per kilogram-year exposure, and take the magnon width to frequency ratio $\gamma/\omega$ to be $10^{-2}$ (solid) or $10^{-5}$ (dashed).
\label{fig:magreach}
}
\end{figure}

To compute the magnon excitation rate, we substitute the coupling $\vect{f}_j$ in Eq.~\eqref{eq:magWcoupling}, into the rate formula Eq.~\eqref{eq:Rmag}. In Sec.~\ref{sec:selection}, we discussed three strategies to alleviate the suppression of axion-magnon couplings due to selection rules: external magnetic fields, anisotropic interactions, and nondegenerate $g$-factors. In this subsection, we show the projected reach for each of these strategies. The results are summarized in Fig.~\ref{fig:magreach}, assuming 3 single magnon events per kilogram-year exposure. Absent a detailed study of anharmonic magnon interactions, we take the resonance widths to be a free parameter, and show results for $\gamma / \omega = 10^{-2}$ and $10^{-5}$, consistent with measured phonon width values on the high end and YIG's Kittel magnon width on the low end. We see that, on resonance, all methods could reach axion-electron couplings predicted by QCD axion models. In the following, we expand on the calculation for each strategy.

\setcounter{paragraph}{0}
\paragraph{External magnetic field.}

The idea of using an external magnetic field to lift the gapless mode is the one adopted in the QUAX experiment~\cite{Ruoso:2015ytk,Barbieri2017a,Crescini:2018qrz}. In Ref.~\cite{Barbieri2017a}, a classical calculation was used to estimate the axion absorption rate. Our formalism allows to compute the same rate in the quantum regime, and agrees with the recent computation carried out in Ref.~\cite{Chigusa2020a}. The projected reach we obtain for a YIG sample in a $1\T$ field is shown in Fig.~\ref{fig:magreach}, where the resonance is at 0.12\,meV (see Eq.~\eqref{eq:omegaL}). For comparison, we also overlay the projection in Ref.~\cite{Chigusa2020a} based on scanning the resonance frequency by changing the magnetic field, with a total integration time of 10 years.

\paragraph{Anisotropic interactions.}

As another way to lift the gapless magnon modes, so that the absorption kinematics are satisfied, we may use materials with anisotropic exchange couplings. As an example, we consider \ce{NiPS3}, which has a layered crystal structure~\cite{Wildes2015a}. The magnetic ions are spin-1 Ni$^{2+}$. Following Ref.~\cite{Kim_2019}, we model the system as having intralayer anisotropic exchange couplings up to third nearest neighbors, as well as single-ion anisotropies. All four magnon branches are gapped, two of which are found to have nonzero couplings to the axion wind. These correspond to the resonances at 12\,meV and 44\,meV in Fig.~\ref{fig:magreach}. 

\paragraph{Nondegenerate $g$-factors.}

Finally, we consider coupling the axion to gapped magnon modes in the presence of nondegenerate $g$-factors. We are not aware of a well-characterized material with nondegenerate $g$-factors so, as a proof of principle, we entertain a few toy models, where a nondegenerate $\ell$ component is added to the effective spins $S$ in YIG. In reality, all the magnetic ions Fe$^{3+}$ in YIG have $(\ell,\, s,\, S) = (0,\, 5/2,\, 5/2)$; the orbital angular momenta of 3d electrons are quenched. In Fig.~\ref{fig:magreach}, we show the reach for two toy models, with either the octahedral sites or the tetrahedral sites modified to have $(\ell,\, s,\, S) = (1,\, 5/2,\, 7/2)$. In each case, only one of the 19 gapped magnon modes, at 7\,meV and 76\,meV respectively, is found to contribute to axion absorption. This is because, to preserve the lattice symmetries, we have modified all the effective spin compositions on tetrahedral or octahedral sites in the same way. The sole gapped mode that couples to axion DM corresponds to out-of-phase precessions of the tetrahedral and octahedral spins.


\section{Conclusions}
\label{sec:conclusions}

In this paper we showed multiple ways collective excitations in crystal targets are, in principle, sensitive to QCD axion DM. Specifically, we identified two novel detection possibilities: axion-induced phonon polariton excitation in an external magnetic field, and axion-induced magnon excitation in the absence of external fields. These paths are complementary as each probes a different axion coupling, $g_{a \gamma \gamma}$ and $g_{aee}$ in the phonon case and magnon case respectively.

In the phonon polariton case, we considered several example targets -- \ce{Al2O3}, \ce{CaWO4}, GaAs and \ce{SiO2} -- and showed that on resonance, per kilogram-year exposure, they can reach $g_{a \gamma \gamma} \sim 10^{-12} \text{ GeV}^{-1}$, as shown in Figs.~\ref{fig:phoreach1} and \ref{fig:phoreach2}. This outperforms the leading constraint in this mass window from stellar cooling, and reaches below the QCD axion band. Carefully choosing a set of target materials with different phonon frequencies is key to covering a broad range of axion masses.

Previous proposals for probing $g_{aee}$ via absorption onto magnons, which underlies the QUAX experiment, considered targets in an external magnetic field, which would lift the lowest magnon mode and kinematically allow sub-meV axion absorption~\cite{Ruoso:2015ytk,Barbieri2017a,Crescini:2018qrz,Flower:2018qgb,Chigusa2020a}. In contrast, we focused on the $\OO(1$-$100)$\,meV axion mass window, and showed that without an external magnetic field, materials with anisotropic exchange interactions, \textit{e.g.}\ \ce{NiPS3}, and materials with nondegenerate $g$-factors can host gapped magnons coupling to the axion. On resonance and with kilogram-year exposure, they can have sensitivity to the DFSZ model and down to $g_{aee}\sim10^{-15}$, shown in Fig.~\ref{fig:magreach}.

Realizing the exciting potential of discovering $\OO(1$-$100)$\,meV axion DM via phonons and magnons hinges upon the ongoing effort to achieve low threshold single quanta detection. One possibility is to evaporate helium atoms from phonon interactions at the surface of the crystal and then detect the evaporated helium atoms in a region separated from the magnetic field region; R\&D is underway for this direction.
Another route is to read out photons produced from the decay of a phonon polariton. Single photon detectors ({\em e.g.}\ superconducting nanowires) may operate in a field-free region, away from the crystal target and connected to it via a waveguide.
Finally, magnons are read out in a resonant cavity in the QUAX setup in the classical regime~\cite{Ruoso:2015ytk,Barbieri2017a,Crescini:2018qrz}. Work is underway to detect single magnons in YIG by coupling cavity modes to a superconducting qubit~\cite{LachanceQuirion2019a}, though as in other resonant cavity searches, axion masses are best produced near the inverse cavity size; for larger axion masses the virtual cavity modes will be off-shell and readout efficiency is suppressed. On the materials side, we would like to make more accurate predictions for the detection rates via an improved understanding of phonon and magnon resonance lineshapes, and explore the possibility of scanning the resonance frequencies by engineering material properties, in order to fully exploit the discovery potential of an axion DM search experiment based on phonon and magnon excitations.


\acknowledgments

We thank Rana Adhikari, Maurice Garcia-Sciveres, Sin\'ead Griffin, Thomas Harrelson, David Hsieh, Stephen Lyon, Matt Pyle and Thomas Schenkel for discussions. A.M., T.T., Z.Z.\ and K.Z.\ are supported by the Quantum Information Science Enabled Discovery (QuantISED) for High Energy Physics (KA2401032). Z.Z.'s work was also supported in part by the NSF grant PHY-1638509.


\appendix


\section{Photon-phonon mixing}
\label{app:polariton}

As outlined in Sec.~\ref{sec:formalism-pho}, mixing between long-wavelength photon and phonon states needs to be taken into account when computing DM absorption rates in a polar crystal. Our starting point is the Lagrangian for an ionic lattice coupling to electromagnetism:
\be
L = \sum_{lj} \biggl(\frac{1}{2}\, m_j \dot{\vect{u}}_{lj}^2 - \frac{1}{2}\sum_{l'j'} \vect{u}_{lj} \cdot \tens{V}_{lj,l'j'}^{(2)} \cdot \vect{u}_{l'j'} 
+e\, \vect{E}(\vect{x}_{lj}^0)\cdot\tens{Z}^*_j\cdot\vect{u}_{lj}\biggr) 
+\int d^3x \biggl( -\frac{1}{4} F^{\mu\nu} F_{\mu\nu} +\frac{1}{2} A_\mu \Pi^{\mu\nu} A_\nu \biggr) .
\label{eq:Lpol}
\ee
In the long wavelength limit, the leading electromagnetic coupling is via the electric dipole, as shown in the third term. Plugging in $\vect{E} = -\nabla A_0 -\dot{\vect{A}}$ and integrate by parts, we can write it in the familiar form of $-\int d^3x \,J^\mu A_\mu = \int d^3x (-\rho A_0 +\vect{J}\cdot\vect{A})$, with
\begin{equation}
\rho(\vect{x}) = -e\sum_{lj} \bigl(\nabla\delta^{(3)}(\vect{x}-\vect{x}_{lj}^0)\bigr) \cdot\tens{Z}^*_j \cdot \vect{u}_{lj} \,,\qquad 
\vect{J}(\vect{x}) = e\sum_{lj}\tens{Z}^*_j\cdot \dot{\vect{u}}_{lj} \,\delta^{(3)}(\vect{x}-\vect{x}_{lj}^0)\,.
\end{equation}
when expanded to linear order in $\vect{u}$. The last term in Eq.~\eqref{eq:Lpol} results from integrating out electron response. As explained in detail in Ref.~\cite{Trickle:2019nya}, the photon self-energy $\Pi^{\mu\nu}$ can be related to the dielectric tensor of the medium, $\vect{\varepsilon}$ (taken to be the electronic contribution, usually denoted by $\vect{\varepsilon}_\infty$, in the present case), and the photon Lagrangian can be written as
\begin{align}
-\frac{1}{4} F^{\mu\nu} F_{\mu\nu} +\frac{1}{2} A_\mu \Pi^{\mu\nu} A_\nu & =
\frac{1}{2} \dot{\vect{A}} \cdot \vect{\varepsilon} \cdot \dot{\vect{A}} - \frac{1}{2}A_0 \left( \nabla \cdot \vect{\varepsilon} \cdot \nabla \right) A_0 + \dot{A}_0 (\nabla \cdot \vect{\varepsilon} \cdot \vect{A})  -\frac{1}{2} \bigl(\partial_i A^j\bigr)^2 +\frac{1}{2} (\nabla\cdot\vect{A})^2\,.
\end{align}
It is convenient to choose a generalized Coulomb gauge, $\nabla \cdot \vect{\varepsilon} \cdot \vect{A}=0$, and since the $A_0$ field is non-dynamical, it can be immediately integrated out. We thus obtain
\beq
L &=& \sum_{lj} \frac{1}{2}\, m_j \dot{\vect{u}}_{lj}^2 - \frac{1}{2}\sum_{ll'jj'} \vect{u}_{lj} \cdot \tens{V}_{ll'jj'} \cdot \vect{u}_{l'j'} +e\,\vect{A}(\vect{x}_{lj}^0) \cdot \tens{Z}^*_j \cdot \dot{\vect{u}}_{lj}  \nonumber \\
&& +\int d^3x \biggl[ \frac{1}{2} \dot{\vect{A}} \cdot \vect{\varepsilon} \cdot \dot{\vect{A}} - \frac{1}{2} \bigl(\partial_i A^j\bigr)^2 + \frac{1}{2} (\nabla\cdot\vect{A})^2 + \frac{1}{2}\, \rho\frac{1}{\nabla \cdot \vect{\varepsilon} \cdot \nabla} \rho \biggr].
\eeq

To derive the Hamiltonian, we note the canonical momenta are:
\begin{equation}
\vect{p}_{lj} = \frac{\partial L}{\partial \dot{\vect{u}}_{lj}} = m_j \dot{\vect{u}}_{lj} +e\, \vect{A}(\vect{x}_{lj}^0) \cdot \tens{Z}^*_j \,,\qquad
\vect{P} (\vect{x}) = \frac{\partial{\cal L}}{\partial \dot{\vect{A}}(\vect{x})} = \vect{\varepsilon} \cdot \dot{\vect{A}}(\vect{x})\,.
\end{equation}
Therefore,
\be
H = \sum_{lj} \vect{p}_{lj}\cdot \dot{\vect{u}}_{lj} +\int d^3x\,\vect{P}\cdot\dot{\vect{A}} -L
= H_\text{ph} + H_\text{Coulomb} + H_\text{EM} + H_\text{mix}\,,
\ee
where
\begin{eqnarray}
H_\text{ph} &=& \sum_{lj} \frac{\vect{p}_{lj}^2}{2m_j} + \frac{1}{2} \sum_{ll'jj'} \vect{u}_{lj} \cdot \tens{V}_{ll'jj'} \cdot \vect{u}_{l'j'}\,, \\
H_\text{Coulomb} &=& -\frac{1}{2} \int d^3x \,\,\rho\,\frac{1}{\nabla \cdot \vect{\varepsilon} \cdot \nabla}\,\rho \,,\\
H_\text{EM} &=& \int d^3x\, \biggl[ \vect{P} \cdot \vect{\varepsilon}^{-1} \cdot \vect{P} + \frac{1}{2} \bigl(\partial_i A^j\bigr)^2 - \frac{1}{2} (\nabla\cdot\vect{A})^2 +\frac{e^2}{2} \sum_{lj} \frac{1}{m_j} \delta^{(3)}(\vect{x}-\vect{x}_{lj}^0) \bigl( \vect{A} \cdot \tens{Z}^*_j\bigr)^2 \biggr]\,,\quad\label{eq:hem} \\
H_\text{mix} &=& -e\sum_{lj} \frac{1}{m_j}\, \vect{A}(\vect{x}_{lj}^0) \cdot \tens{Z}^*_j \cdot \vect{p}_{lj}\,.
\end{eqnarray}
Note that while we have written $\vect{A}$ and $\vect{P}$ as 3-vectors, they are implicitly assumed to satisfy the gauge condition. This means that, in Eq.~\eqref{eq:hem}, $\vect{\varepsilon}$ should be projected onto the subspace satisfying the gauge condition before the inverse is taken.

We now consider the four terms in turn. First, as mentioned in Sec.~\ref{sec:formalism-pho}, we use the \texttt{phonopy} code with DFT calculations of the force constants $\tens{V}_{lj,l'j'}^{(2)}$ to diagonalize the lattice (phonon) Hamiltonian $H_\text{ph}$, giving
\begin{equation}
H_\text{ph} = \sum_{\nu=1}^{3n}\sum_{\vect{k}} \omega_{\nu,\vect{k}} \hat a^\dagger_{\nu,\vect{k}} \hat a_{\nu,\vect{k}} \,.
\label{eq:Hph}
\end{equation}

Second, the Coulomb term, $H_\text{Coulomb}$, becomes more transparent when written as a momentum space integral:
\begin{equation}
\label{eq:HCoulomb1}
H_\text{Coulomb} = \frac{1}{2}\int\frac{d^3k}{(2\pi)^3} \frac{|\tilde\rho(\vect{k})|^2}{ \vect{k} \cdot \vect{\varepsilon} \cdot \vect{k}}\,,
\end{equation}
with the charge density
\begin{equation}
\tilde \rho(\vect{k}) = \int d^3x \,e^{-i\vect{k}\cdot\vect{x}} \rho(\vect{x})
= -ie\sum_{lj} \vect{k}\cdot\tens{Z}^*_j \cdot\vect{u}_{lj} \,e^{-i\vect{k}\cdot\vect{x}_{lj}^0} \,.
\end{equation}
Expanding $\vect{u}_{lj}$ in $(\hat a_{\nu,\vect{k}'}+\hat a^\dagger_{\nu,-\vect{k}'})$ as in Eq.~\eqref{eq:ulj} and summing over $l$ picks out the $\vect{k}'=\vect{k}$ modes. With the momentum integral discretized, $\int\frac{d^3k}{(2\pi)^3}\to \frac{1}{N\Omega}\sum_{\vect{k}}$, we find
\begin{equation}
H_\text{Coulomb} = \frac{e^2}{4\Omega} \sum_{\nu,\nu',\vect{k}} \frac{1}{\sqrt{\omega_{\nu',\vect{k}}\omega_{\nu,\vect{k}}}} \frac{(\vect{k}\cdot\vect{\xi}^*_{\nu',\vect{k}})(\vect{k}\cdot\vect{\xi}_{\nu,\vect{k}})}{\vect{k} \cdot \vect{\varepsilon} \cdot \vect{k}} \bigl( \hat a_{\nu',-\vect{k}} +\hat a^\dagger_{\nu',\vect{k}}\bigr) \bigl( \hat a_{\nu,\vect{k}} +\hat a^\dagger_{\nu,-\vect{k}}\bigr)\,,
\label{eq:HCoulomb2}
\end{equation}
where
\be
\vect{\xi}_{\nu,\vect{k}} \equiv \sum_j \frac{1}{\sqrt{m_j}}\, \tens{Z}^*_j \cdot \vect{\epsilon}_{\nu,\vect{k},j}\,.
\ee
As a technical note, while there is an option in \texttt{phonopy} (non-analytic correction) to also include $H_\text{Coulomb}$ in the diagonalization calculation, it seems to work only at $k\gtrsim\omega$. Therefore, in our calculation, we use \texttt{phonopy} to diagonalize only $H_\text{ph}$, and include $H_\text{Coulomb}$ separately.

Next, we also write the photon Hamiltonian $H_\text{EM}$ in momentum space:
\begin{equation}
H_\text{EM} = \int \frac{d^3k}{(2\pi)^3} \biggl[ \frac{1}{2} \vect{\tilde{P}}(\vect{k})^* \cdot \vect{\varepsilon}^{-1} \cdot \vect{\tilde{P}}(\vect{k}) +\frac{1}{2}\, \vect{\tilde{A}}(\vect{k})^* \cdot \tens{K^2} \cdot \vect{\tilde{A}}(\vect{k}) \biggr]\,,
\end{equation}
where
\begin{equation}
\label{eq:K}
\tens{K^2} = k^2 \mathbb{1} +\frac{e^2}{\Omega} \sum_j \frac{\tens{Z}^{*}_j \tens{Z}^{*T}_j}{m_j} - \vect{k}\vect{k}\,.
\end{equation}
We decompose the photon field $\vect{\tilde{A}}$ into two orthogonal linear polarizations $\vect{e}_{1,\vect{k}}\perp \vect{e}_{2,\vect{k}}$ which satisfy the gauge condition, $\vect{k} \cdot \vect{\varepsilon} \cdot \vect{e}_{\lambda, \vect{k}} = 0$ ($\lambda = 1,2$). We choose the basis in which the projection of $\vect{\varepsilon}$ onto the two-dimensional subspace, $\vect{e}_{\lambda'}\cdot\vect{\varepsilon}\cdot\vect{e}_\lambda$, is diagonal, with eigenvalues $\varepsilon_1$, $\varepsilon_2$. Denote the projection of $\tens{K^2}$ in this basis by
\begin{align}
\vect{e}_{\lambda',\vect{k}}^* \cdot \tens{K^2} \cdot \vect{e}_{\lambda,\vect{k}}& \equiv K^2_{\lambda'\lambda}\,.
\end{align}
We introduce phonon creation and annihilation operators $\hat b^\dagger_{\lambda,\vect{k}}$, $\hat b_{\lambda,\vect{k}}$ satisfying the usual (discretized) commutation relations, $[\hat b_{\lambda,\vect{k}},\,\hat b^\dagger_{\lambda',\vect{k}'}] =\delta_{\lambda,\lambda'}\delta_{\vect{k},\vect{k}'}$, etc., based on the diagonal piece of the Hamiltonian:
\begin{eqnarray}
\vect{\tilde{A}}(\vect{k}) &=& \sqrt{N\Omega} \sum_\lambda \frac{1}{\sqrt{2\,\varepsilon_\lambda^{1/2} K_{\lambda\lambda}}} \,\bigl( \hat b_{\lambda,\vect{k}} +\hat b^\dagger_{\lambda,-\vect{k}} \bigr) \,\vect{e}_{\lambda,\vect{k}} \,,\label{eq:Ak} \\
\vect{\tilde{P}}(\vect{k}) &=& \sqrt{N\Omega} \sum_\lambda \sqrt{\frac{\varepsilon_\lambda^{1/2} K_{\lambda\lambda}}{2}}\, \frac{1}{i} \bigl( \hat b_{\lambda,\vect{k}} -\hat b^\dagger_{\lambda,-\vect{k}} \bigr) \,\vect{e}_{\lambda,\vect{k}} \,.
\end{eqnarray}
The photon Hamiltonian then becomes
\begin{equation}
H_\text{EM} = \sum_{\vect{k}} \biggl[ \sum_{\lambda=1}^2 \frac{K_{\lambda\lambda}}{\sqrt{\varepsilon_\lambda}} \,\hat b_{\lambda,\vect{k}}^\dagger \hat b_{\lambda,\vect{k}}\, + \frac{K_{12}^2}{2\sqrt{\sqrt{\varepsilon_1 \varepsilon_2} K_{11}K_{22}}} \bigl( \hat b_{1,-\vect{k}} +\hat b^\dagger_{1,\vect{k}} \bigr)\bigl( \hat b_{2,\vect{k}} +\hat b^\dagger_{2,-\vect{k}} \bigr) \biggr]\,,
\label{eq:HEM}
\end{equation}
where we have used $K_{12}=K_{21}$.

Finally, the photon-phonon mixing term $H_\text{mix}$ can be written in terms of the creation and annihilation operators according to Eqs.~\eqref{eq:plj} and \eqref{eq:Ak}:
\begin{eqnarray}
H_\text{mix} &=& -\frac{e}{N\Omega}\sum_{\vect{k}} \sum_{lj} \frac{e^{i\vect{k}\cdot\vect{x}_{lj}^0}}{m_j}  \vect{\tilde{A}}(\vect{k})\cdot \tens{Z}^*_j \cdot \vect{p}_{lj}  \nonumber\\
&=& \frac{ie}{2\sqrt{\Omega}} \sum_{\vect{k}} \sum_{\nu=1}^{3n} \sum_{\lambda=1}^2 \sqrt{\frac{\omega_{\nu,\vect{k}}}{\varepsilon_\lambda^{1/2} K_{\lambda\lambda}}}\, \bigl( \vect{e}_{\lambda,-\vect{k}} \cdot \vect{\xi}_{\nu,\vect{k}} \bigr) \bigl( \hat a_{\nu,\vect{k}} -\hat a^\dagger_{\nu,-\vect{k}} \bigr) \bigl( \hat b_{\lambda,-\vect{k}} +\hat b^\dagger_{\lambda,\vect{k}} \bigr) \,.
\label{eq:Hmix}
\end{eqnarray}

The total quadratic Hamiltonian, given by the sum of Eqs.~\eqref{eq:Hph}, Eqs.~\eqref{eq:HCoulomb2}, Eqs.~\eqref{eq:HEM} and Eqs.~\eqref{eq:Hmix}, involves the $3n$ phonon and 2 photon creation/annihilation operators, $\hat a^{(\dagger)}_{\nu=(1,\dots,3n),\vect{k}}$, $\hat b^{(\dagger)}_{\lambda=(1,2),\vect{k}}$. For simplicity, let us write $\hat a^{(\dagger)}_{\nu=(3n+1,3n+2),\vect{k}} \equiv \hat b^{(\dagger)}_{\lambda=(1,2),\vect{k}}$. The quadratic Hamiltonian then has the form:
\beq\label{eq:polarham}
\hat{H}=\sum_{\vect{k}\in 1{\textrm BZ}}\mathbb{a}^\dagger_{\vect{k}}\cdot\mathbb{h}_{\vect{k}}\cdot\mathbb{a}_{\vect{k}}\qquad \textrm{with}\qquad \mathbb{a}_{\vect{k}}=\left[\hat a_{1,\vect{k}}, \ldots,\,\hat a_{3n+2,\vect{k}},\hat a^\dagger_{1,\vect{-k}},\ldots,\,\hat a^\dagger_{3n+2,\vect{-k}}\right]^{\mathsf{T}}\,.
\eeq
The matrix $\mathbb{h}_{\vect{k}}$ can be written as:
\beq
\mathbb{h}_{\vect{k}}=\left(
\arraycolsep=3pt\def\arraystretch{0.7}
\begin{array}{ccccc?ccccc}
 &&& \rvline &&&&& \rvline &   \\
&\mathbb{A}_{\vect{k}}  && \rvline &  \mathbb{B}^*_{\vect{k}}&&\mathbb{A}_{\vect{k}}  && \rvline &  \mathbb{B}^\dagger_{\vect{k}} \\
 &&& \rvline &&&&& \rvline &   \\
\hline
 & \mathbb{B}^{\mathsf{T}}_{\vect{k}} && \rvline & \mathbb{C}_{\vect{k}}& & -\mathbb{B}^{\mathsf{T}}_{\vect{k}} && \rvline & \mathbb{C}_{\vect{k}}\\
 \thickhline
  &&& \rvline &&&&& \rvline &   \\
&\mathbb{A}_{\vect{k}}  && \rvline &  \mathbb{B}_{-\vect{k}}&&\mathbb{A}_{\vect{k}}  && \rvline &  \mathbb{B}_{-\vect{k}} \\
 &&& \rvline &&&&& \rvline &   \\
\hline
 & \mathbb{B}^{\mathsf{T}}_{\vect{k}} && \rvline & \mathbb{C}_{\vect{k}}&& -\mathbb{B}^{\mathsf{T}}_{\vect{k}} && \rvline & \mathbb{C}_{\vect{k}}
\end{array}\right),
\eeq
where $\mathbb{A}_{\vect{k}}$ is a $3n\times 3n$ matrix given by: 
\beq\label{eq:Amatrix1}
\mathbb{A}_{\vect{k},\nu\nu'}=\frac{1}{2}\omega_{\nu}\delta_{\nu\nu'}+\frac{e^2}{4\Omega\sqrt{\omega_{\nu,\bk}\omega_{\nu',\bk}}}\frac{(\bk\cdot\vect{\xi}_{\nu,\bk}^*)(\bk\cdot\vect{\xi}_{\nu',\bk})}{\vect{k} \cdot \vect{\varepsilon} \cdot \vect{k}}\,,
\eeq
while $\mathbb{B}_{\vect{k}}$ is a $3n\times2$ with the following structure 
\beq\label{eq:Bmatrix1}
\mathbb{B}_{\vect{k},\nu\nu'}=-\sum_{\lambda=1}^2\frac{ie}{4\sqrt\Omega}\sqrt{\frac{\omega_{\nu,\bk}}{\varepsilon_{\nu'}^{1/2}K_{\nu'\nu'}}}\left(\vect{\xi}_{\nu,\vect{k}}\cdot\vect{e}_{\nu',-\vect{k}}\right)\,,
\eeq
and $\mathbb{C}_{\vect{k},\nu\nu'}$ is a $2\times 2$ matrix given by 
\beq\label{eq:Cmatrix1}
\mathbb{C}_{\vect{k},\nu\nu'}=\frac{K_{\nu\nu'}^2}{4\sqrt{\sqrt{\varepsilon_\nu \varepsilon_{\nu'}} K_{\nu\nu}K_{\nu'\nu'}}}\,.
\eeq
We review a general algorithm for diagonalizing such Hamiltonians in Appendix~\ref{app:diag}.


\section{General form of the magnon Hamiltonian}
\label{app:magnon}
In this appendix, by following Ref.~\cite{Toth-Lake}, we review the procedure to derive the quadratic Hamiltonian describing small fluctuations around the ground state of the spin lattice described by \Eq{eq:SpinHam}.\footnote{We adopt a different phase convention compared to Ref.~\cite{Toth-Lake}, by using $e^{i\vect{k}\cdot\vect{x}_{lj}}$ as oppose to $e^{i\vect{k}\cdot\vect{x}_{l}}$, in the Fourier transform Eq.~\eqref{eq:Fourier}.} The results we review will apply to both commensurate and single-$Q$ incommensurate materials (such as \ce{Ba3NbFe3Si2O14} discussed in Sec.~\ref{sec:selection}). We conclude the appendix giving the generalization of the rate in \Eq{eq:Rmag} for the case of incommensurate materials. 

To include in our discussion the case of single-$Q$ incommensurate materials, we need to generalize \Eq{eq:SpinRot} to 
\beq\label{eq:spinrot}
\vect{S}_{lj} = \tens{R}'(\vect{x}_{lj}) \cdot \tens{R}_j \cdot \vect{S}_{lj}' \,,
\eeq
where the additional rotation $\vect{R}'(\vect{x}_{lj})$ brings to a reference frame where the spin orientation looks the same in all the unit cells (for commensurate ordered materials, $\vect{R}'=\tens{1}$). Following the Holstein-Primakoff transformation, Eq.~\eqref{eq:HolPri}, we we can rewrite \Eq{eq:spinrot} as, at leading order,
\beq\label{eq:Spinop}
\vect{S}_{lj}=\tens{R}'(\vect{x}_{lj})\left[\sqrt{\frac{S_j}{2}}\left(\vect{r}^*_j\hat a_{lj}+\vect{r}_j\hat a_{lj}^\dagger\right)+\vect{t}_j\left(S_j-\hat a^\dagger_{lj}\hat a_{lj}\right)\right],
\eeq
where $\vect{t}_j$ is a unit vector pointing along the direction of the $j$-th spin in the rotating frame (\emph{i.e.} the reference frame defined by the rotation $\tens{R}'$), while $\vect{r}_j$ and $\vect{r}_j^*$ span an orthogonal coordinate system. These vectors are related to the components of the matrix $\tens{R}_j$ by 
\beq
\vect{r}_j^\alpha=\tens{R}_j^{\alpha1}+i\tens{R}_j^{\alpha2}\qquad\qquad \vect{t}_j^\alpha=\tens{R}_j^{\alpha3}\,.
\eeq 
By substituting \Eq{eq:Spinop} into \Eq{eq:SpinHam}, and going to momentum space, we obtain the following expression for the quadratic part of the Hamiltonian 
\beq\label{eq:quadMagHam}
\hat{H}=\sum_{\vect{k}\in 1{\textrm BZ}}\mathbb{a}^\dagger_{\vect{k}}\cdot\mathbb{h}_{\vect{k}}\cdot\mathbb{a}_{\vect{k}}\qquad \textrm{with}\qquad \mathbb{a}_{\vect{k}}=\left[\hat a_{1,\vect{k}}, \ldots,\,\hat a_{n,\vect{k}},\hat a^\dagger_{1,\vect{-k}},\ldots,\,\hat a^\dagger_{n,\vect{-k}}\right]^{\mathsf{T}}\,.
\eeq
The matrix $\mathbb{h}_{\vect{k}}$ can be written in terms of $n\times n$ sub-matrices as:
\beq
\mathbb{h}_{\vect{k}}=\left(\begin{array}{cc}
\mathbb{A}_{\vect{k}}-\mathbb{C}&\mathbb{B}_{\vect{k}}\\
\mathbb{B}^\dagger_{\vect{k}} & \mathbb{A}^*_{-\vect{k}}-\mathbb{C}\end{array}\right)
\eeq
where, by defining $\tens{J}'_{lj,l'j'}=\tens{R}'(\vect{x}_{lj})^{\mathsf{T}}\cdot\tens{J}_{lj,l'j'}\cdot\tens{R}'(\vect{x}_{l'j'})$, we have
\beq
\begin{aligned}
\mathbb{A}_{\vect{k},jj'}&=\mathbb{A}^\dagger_{\vect{k},jj'}=\frac{\sqrt{S_jS_{j'}}}{2}\vect{r}_j^{\mathsf{T}}\cdot\tens{J}'_{-\vect{k},jj'}\cdot\vect{r}_{j'}^*-\frac{1}{2}\mu_{\textrm{B}}\delta_{jj'}\,\vect{B}^{\mathsf{T}}\cdot g_{j}\cdot\vect{t}_j\\
\mathbb{B}_{\vect{k},jj'}&=\mathbb{B}^*_{\vect{k},jj'}=\frac{\sqrt{S_jS_{j'}}}{2}\vect{r}_j^{\mathsf{T}}\cdot\tens{J}'_{-\vect{k},jj'}\cdot\vect{r}_{j'}\\
\mathbb{C}_{\vect{k},jj'}&=\delta_{jj'}\sum_lS_l\,\vect{t}_j^{\mathsf{T}}\cdot\tens{J}'_{jl}(0)\cdot\vect{t}_l
\end{aligned}\,.
\eeq
The procedure to diagonalize Hamiltonian of this kind is reviewed in Appendix~\ref{app:diag}.

We conclude by generalizing to the incommensurate case the formula for magnon production given in \Eq{eq:Rmag}.
We start by writing the rotation matrix $\vect{R}'(\vect{x}_{lj})$ in terms of the propagation vector which characterizes the incommensurate order $\vect{Q} = (\tau_1,\tau_2,\tau_3)$:
\begin{equation}
\tens{R}'(\vect{x}_{lj}) = \mathcal{R} (\vect{n}\,|\, \vect{Q}\cdot\vect{x}_{lj})\,,
\end{equation}
where $\mathcal{R} (\vect{n}\,|\, \varphi)$ is the rotation matrix around the unit vector $\vect{n}$ by an angle $\varphi$:
\begin{eqnarray}
&&
\mathcal{R} (\vect{n}\,|\, \varphi) = \text{Re} \bigl[ e^{i\varphi} (\vect{1}-i\tens{n}_\times -\vect{n}\vect{n}^T) \bigr] +\vect{n}\vect{n}^T
= \tens{R}'_0 +\tens{R}'_+ e^{i\varphi} +\tens{R}'_- e^{-i\varphi} \\
&&
\tens{R}'_0 = \vect{n}\vect{n}^T\,,\qquad
\tens{R}'_\pm = \frac{1}{2} (\tens{1}\mp i\tens{n}_\times -\vect{n}\vect{n}^T)\,,\qquad
\tens{n}_\times = \left(
\begin{matrix}
0 & -n_z & n_y \\
n_z & 0 & -n_x \\
-n_y & n_x & 0
\end{matrix}
\right)\,,
\end{eqnarray}
noting that $\tens{n}_\times \cdot \vect{v}=\vect{n}\times\vect{v}$ for any vector $\vect{v}$. Then, by using \Eq{eq:spinrot} and the interaction given by Eq.~\eqref{eq:fj}, we find
\begin{eqnarray}
\langle \nu, \vect{k} | \hat{\delta H_0} | 0 \rangle &=& \sum_{lj} \sqrt{\frac{S_j}{2N}}\, e^{-i\vect{k}\cdot\vect{x}_{lj}} \,\vect{f}_j \cdot \tens{R}'(\vect{x}_{lj}) \cdot \bigl(\mathbb{V}_{j\nu,-\vect{k}} \vect{r}_j^* +\mathbb{U}^*_{j\nu,\vect{k}}\vect{r}_j\bigr) \nonumber\\
&=& \sqrt{\frac{N}{2}} \sum_j \sqrt{S_j}\, \vect{f}_j \cdot \bigl(\delta_{\vect{k},\vect{0}} \tens{R}'_0 + \delta_{\vect{k},\vect{Q}}\,\tens{R}'_+ + \delta_{\vect{k},-\vect{Q}}\,\tens{R}'_- \bigr) \cdot \bigl(\mathbb{V}_{j\nu,-\vect{k}} \vect{r}_j^* +\mathbb{U}^*_{j\nu,\vect{k}}\vect{r}_j\bigr)\,,
\end{eqnarray}
from which we can obtain the generalized expression for the rate:
\begin{equation}
R_{\vect{Q} \neq 0} 
= \frac{2\omega}{m_\text{cell}}\sum_{\nu = 1}^n \sum_{\lambda = -1}^1 \,\frac{\omega_{\nu,\lambda\vect{Q}}\,\gamma_{\nu,\lambda\vect{Q}}}{(\omega^2-\omega_{\nu,\lambda\vect{Q}}^2)^2+(\omega\, \gamma_{\nu,\lambda\vect{Q}})^2}\, \biggl|\sum_j \sqrt{S_j}\, \vect{f}_j^T\,\tens{R}'_\lambda \,\bigl(\mathbb{V}_{j\nu,-\lambda\vect{Q}} \vect{r}_j^* +\mathbb{U}^*_{j\nu,\lambda\vect{Q}}\vect{r}_j\bigr) \biggr|^2 \,.
\label{eq:Rmag-i}
\end{equation}


\section{Diagonalization of quadratic Hamiltonians}
\label{app:diag}
In this appendix, by closely following Ref.~\cite{Toth-Lake,Colpa1978a}, we review the procedure to diagonalize quadratic Hamiltonians which have the form of \Eq{eq:polarham} and \Eq{eq:quadMagHam}. The goal of the procedure is to find a homogeneous linear transformation 
\beq
\mathbb{a}_{\vect{k}}=\mathbb{T}_{\vect{k}}\cdot\mathbb{a}_{\vect{k}}'\equiv\mathbb{T}_{\vect{k}}\cdot\left[\hat a_{1,\vect{k}}', \ldots,\,\hat a_{n,\vect{k}}',\hat a'^{\dagger}_{1,-\vect{k}},\ldots,\,\hat a'^{\dagger}_{n,-\vect{k}}\right]^{\mathsf{T}}\,,
\eeq
such that the operators $\hat a'_{\nu,\vect{k}}$, $\hat a_{\nu,\vect{k}}^{'\dagger}$ satisfy the canonical commutation relations
\beq\label{eq:ComRel}
[\mathbb{a}_{\vect{k}}',\mathbb{a}_{\vect{k}}'^{\dagger}]=\left(\begin{array}{cc}\mathbb{1} & 0\\0 & -\mathbb{1}\end{array}\right)\equiv \mathbb{g}\,,
\eeq
and that rewrites the quadratic Hamiltonian as 
\beq\label{eq:DiagHam}
\hat{H}&=\sum_{\vect{k}}\mathbb{a}^\dagger_{\vect{k}}\mathbb{h}_{\vect{k}}\mathbb{a}_{\vect{k}}=\sum_{\vect{k}} \mathbb{a}'^{\dagger}_{\vect{k}}\cdot\mathbb{E}\cdot\mathbb{a}'_{\vect{k}}\,,
\eeq
where 
\beq \label{eq:Emat}
\mathbb{E}_{\vect{k}}\equiv \mathbb{T}^\dagger_{\vect{k}}\cdot\mathbb{h}_{\vect{k}}\cdot\mathbb{T}_{\vect{k}}=\frac{1}{2}\,\textrm{diag}\left(\omega_{1,\vect{k}}\ldots\omega_{n,\vect{k}},\omega_{1,\vect{k}}\ldots\omega_{n,\vect{k}}\right)\qquad\textrm{with}\qquad\omega_{\nu,\vect{k}}>0\,.
\eeq
By using the commutation relations in \Eq{eq:ComRel}, it can be easily shown that, up to constant terms, \Eq{eq:DiagHam} is equivalent to \Eq{eq:diagHamPho} and \Eq{eq:MagHamDia} with the $\mathbb{U}$ and $\mathbb{V}$ matrices implicitly defined as 
\beq
\mathbb{T}_{\vect{k}} = 
\left(\begin{matrix}
\mathbb{U}_{j\nu,\vect{k}} & \mathbb{V}_{j\nu,\vect{k}} \\
\mathbb{V}_{j\nu,-\vect{k}}^* & \mathbb{U}_{j\nu,-\vect{k}}^*
\end{matrix}\right)
\label{eq:diag}\,.
\eeq

Such diagonalization procedure (usually called para-unitary diagonalization) can be achieved if $\mathbb{h}(\vect{k})$ is positive definite. If the spectrum contains zero energy modes, the $\mathbb{h}(\vect{k})$ matrix will be positive \emph{semidefinite} and such para-unitary diagonalization may not exist. This problem can be cured by adding a small $\epsilon$ value to the diagonal components of $\mathbb{h}(\vect{k})$~\cite{Colpa1978a}. This introduces a small and negligible shift in the spectrum but makes $\mathbb{h}(\vect{k})$ positive definite and allows for a para-unitary diagonalization. Ref.~\cite{Colpa1978a} provides a simple three-step algorithm to find the linear transformation $\mathbb{T}$ and the associated eigenvalues:
\begin{itemize}
\item A Cholesky decomposition is applied to find a complex matrix $\mathbb{K}_{\vect{k}}$ such that $\mathbb{h}_{\vect{k}}=\mathbb{K}_{\vect{k}}^\dagger \mathbb{K}_{\vect{k}}$.
\item The eigenvalue problem for the Hermitian matrix $\mathbb{K}_{\vect{k}}\mathbb{g}\mathbb{K}_{\vect{k}}^\dagger$ is solved, and the resulting eigenvalues used to form the columns of the matrix $\mathbb{U}_{\vect{k}}$. The order of the columns is chosen such that the first $N$ elements of the diagonalized matrix $\mathbb{L}=\mathbb{U}^\dagger\mathbb{K}_{\vect{k}} \mathbb{g}\mathbb{K}^\dagger_{\vect{k}}\mathbb{U}$ are positive and the last $N$ negative.
\item Finally, the matrix $\mathbb{E}_{\vect{k}}$ in \Eq{eq:Emat} is simply related to $\mathbb{L}_{\vect{k}}$ by $\mathbb{E}_{\vect{k}} = \mathbb{g}\mathbb{L}_{\vect{k}}$, and the $\mathbb{T}_{\vect{k}}$ matrix is given by $\mathbb{T}_{\vect{k}}=\mathbb{K}_{\vect{k}}^{-1}\mathbb{U}_{\vect{k}}\mathbb{E}_{\vect{k}}^{1/2}$.
\end{itemize}

\bibliography{Biblio}

\providecommand{\href}[2]{#2}\begingroup\raggedright\begin{thebibliography}{10}

\bibitem{Weinberg1975a}
S.~Weinberg, \emph{{The U(1) problem}},
  \href{https://doi.org/10.1103/physrevd.11.3583}{\emph{Physical Review D}
  {\bfseries 11} (jun, 1975) 3583--3593}.

\bibitem{Peccei1977a}
R.~D. Peccei and H.~R. Quinn, \emph{{CP Conservation in the Presence of
  Pseudoparticles}},
  \href{https://doi.org/10.1103/physrevlett.38.1440}{\emph{Physical Review
  Letters} {\bfseries 38} (jun, 1977) 1440--1443}.

\bibitem{Peccei1977b}
R.~D. Peccei and H.~R. Quinn, \emph{{Constraints imposed by CP conservation in
  the presence of pseudoparticles}},
  \href{https://doi.org/10.1103/physrevd.16.1791}{\emph{Physical Review D}
  {\bfseries 16} (sep, 1977) 1791--1797}.

\bibitem{Wilczek1978a}
F.~Wilczek, \emph{{Problem of Strong P and T Invariance in the Presence of
  Instantons}},
  \href{https://doi.org/10.1103/physrevlett.40.279}{\emph{Physical Review
  Letters} {\bfseries 40} (jan, 1978) 279--282}.

\bibitem{Asztalos2010a}
{\scshape ADMX} collaboration, S.~J. Asztalos et~al., \emph{{A SQUID-based
  microwave cavity search for dark-matter axions}},
  \href{https://doi.org/10.1103/PhysRevLett.104.041301}{\emph{Phys. Rev. Lett.}
  {\bfseries 104} (2010) 041301},
  [\href{https://arxiv.org/abs/0910.5914}{{\ttfamily 0910.5914}}].

\bibitem{Du2018a}
{\scshape ADMX} collaboration, N.~Du et~al., \emph{{A Search for Invisible
  Axion Dark Matter with the Axion Dark Matter Experiment}},
  \href{https://doi.org/10.1103/PhysRevLett.120.151301}{\emph{Phys. Rev. Lett.}
  {\bfseries 120} (2018) 151301},
  [\href{https://arxiv.org/abs/1804.05750}{{\ttfamily 1804.05750}}].

\bibitem{Zhong2018a}
{\scshape HAYSTAC} collaboration, L.~Zhong et~al., \emph{{Results from phase 1
  of the HAYSTAC microwave cavity axion experiment}},
  \href{https://doi.org/10.1103/PhysRevD.97.092001}{\emph{Phys. Rev.}
  {\bfseries D97} (2018) 092001},
  [\href{https://arxiv.org/abs/1803.03690}{{\ttfamily 1803.03690}}].

\bibitem{Ouellet:2018beu}
J.~L. Ouellet et~al., \emph{{First Results from ABRACADABRA-10 cm: A Search for
  Sub-$\mu$eV Axion Dark Matter}},
  \href{https://doi.org/10.1103/PhysRevLett.122.121802}{\emph{Phys. Rev. Lett.}
  {\bfseries 122} (2019) 121802},
  [\href{https://arxiv.org/abs/1810.12257}{{\ttfamily 1810.12257}}].

\bibitem{Garcon:2019inh}
A.~Garcon et~al., \emph{{Constraints on bosonic dark matter from ultralow-field
  nuclear magnetic resonance}},
  \href{https://arxiv.org/abs/1902.04644}{{\ttfamily 1902.04644}}.

\bibitem{Anastassopoulos2017a}
{\scshape CAST} collaboration, V.~Anastassopoulos et~al., \emph{{New CAST Limit
  on the Axion-Photon Interaction}},
  \href{https://doi.org/10.1038/nphys4109}{\emph{Nature Phys.} {\bfseries 13}
  (2017) 584--590}, [\href{https://arxiv.org/abs/1705.02290}{{\ttfamily
  1705.02290}}].

\bibitem{TheMADMAXWorkingGroup:2016hpc}
{\scshape MADMAX Working Group} collaboration, A.~Caldwell, G.~Dvali,
  B.~Majorovits, A.~Millar, G.~Raffelt, J.~Redondo et~al., \emph{{Dielectric
  Haloscopes: A New Way to Detect Axion Dark Matter}},
  \href{https://doi.org/10.1103/PhysRevLett.118.091801}{\emph{Phys. Rev. Lett.}
  {\bfseries 118} (2017) 091801},
  [\href{https://arxiv.org/abs/1611.05865}{{\ttfamily 1611.05865}}].

\bibitem{Brun2019a}
{\scshape MADMAX} collaboration, P.~Brun et~al., \emph{{A new experimental
  approach to probe QCD axion dark matter in the mass range above 40 $\mu$eV}},
  \href{https://doi.org/10.1140/epjc/s10052-019-6683-x}{\emph{Eur. Phys. J.}
  {\bfseries C79} (2019) 186},
  [\href{https://arxiv.org/abs/1901.07401}{{\ttfamily 1901.07401}}].

\bibitem{Ruoso:2015ytk}
G.~Ruoso, A.~Lombardi, A.~Ortolan, R.~Pengo, C.~Braggio, G.~Carugno et~al.,
  \emph{{The QUAX proposal: a search of galactic axion with magnetic
  materials}}, \href{https://doi.org/10.1088/1742-6596/718/4/042051}{\emph{J.
  Phys. Conf. Ser.} {\bfseries 718} (2016) 042051},
  [\href{https://arxiv.org/abs/1511.09461}{{\ttfamily 1511.09461}}].

\bibitem{Barbieri2017a}
R.~Barbieri, C.~Braggio, G.~Carugno, C.~S. Gallo, A.~Lombardi, A.~Ortolan
  et~al., \emph{{Searching for galactic axions through magnetized media: the
  QUAX proposal}},
  \href{https://doi.org/10.1016/j.dark.2017.01.003}{\emph{Phys. Dark Univ.}
  {\bfseries 15} (2017) 135--141},
  [\href{https://arxiv.org/abs/1606.02201}{{\ttfamily 1606.02201}}].

\bibitem{Crescini:2018qrz}
N.~Crescini et~al., \emph{{Operation of a ferromagnetic axion haloscope at
  $m_a=58\,\mu$eV}},
  \href{https://doi.org/10.1140/epjc/s10052-018-6163-8}{\emph{Eur. Phys. J. C}
  {\bfseries 78} (2018) 703},
  [\href{https://arxiv.org/abs/1806.00310}{{\ttfamily 1806.00310}}].

\bibitem{DeRocco:2018jwe}
W.~DeRocco and A.~Hook, \emph{{Axion interferometry}},
  \href{https://doi.org/10.1103/PhysRevD.98.035021}{\emph{Phys. Rev. D}
  {\bfseries 98} (2018) 035021},
  [\href{https://arxiv.org/abs/1802.07273}{{\ttfamily 1802.07273}}].

\bibitem{Obata:2018vvr}
I.~Obata, T.~Fujita and Y.~Michimura, \emph{{Optical Ring Cavity Search for
  Axion Dark Matter}},
  \href{https://doi.org/10.1103/PhysRevLett.121.161301}{\emph{Phys. Rev. Lett.}
  {\bfseries 121} (2018) 161301},
  [\href{https://arxiv.org/abs/1805.11753}{{\ttfamily 1805.11753}}].

\bibitem{Liu:2018icu}
H.~Liu, B.~D. Elwood, M.~Evans and J.~Thaler, \emph{{Searching for Axion Dark
  Matter with Birefringent Cavities}},
  \href{https://doi.org/10.1103/PhysRevD.100.023548}{\emph{Phys. Rev. D}
  {\bfseries 100} (2019) 023548},
  [\href{https://arxiv.org/abs/1809.01656}{{\ttfamily 1809.01656}}].

\bibitem{Flower:2018qgb}
G.~Flower, J.~Bourhill, M.~Goryachev and M.~E. Tobar, \emph{{Broadening
  frequency range of a ferromagnetic axion haloscope with strongly coupled
  cavity--magnon polaritons}},
  \href{https://doi.org/10.1016/j.dark.2019.100306}{\emph{Phys. Dark Univ.}
  {\bfseries 25} (2019) 100306},
  [\href{https://arxiv.org/abs/1811.09348}{{\ttfamily 1811.09348}}].

\bibitem{Nagano:2019rbw}
K.~Nagano, T.~Fujita, Y.~Michimura and I.~Obata, \emph{{Axion Dark Matter
  Search with Interferometric Gravitational Wave Detectors}},
  \href{https://doi.org/10.1103/PhysRevLett.123.111301}{\emph{Phys. Rev. Lett.}
  {\bfseries 123} (2019) 111301},
  [\href{https://arxiv.org/abs/1903.02017}{{\ttfamily 1903.02017}}].

\bibitem{Lawson:2019brd}
M.~Lawson, A.~J. Millar, M.~Pancaldi, E.~Vitagliano and F.~Wilczek,
  \emph{{Tunable axion plasma haloscopes}},
  \href{https://doi.org/10.1103/PhysRevLett.123.141802}{\emph{Phys. Rev. Lett.}
  {\bfseries 123} (2019) 141802},
  [\href{https://arxiv.org/abs/1904.11872}{{\ttfamily 1904.11872}}].

\bibitem{Berlin:2019ahk}
A.~Berlin, R.~T. D'Agnolo, S.~A. Ellis, C.~Nantista, J.~Neilson, P.~Schuster
  et~al., \emph{{Axion Dark Matter Detection by Superconducting Resonant
  Frequency Conversion}},  \href{https://arxiv.org/abs/1912.11048}{{\ttfamily
  1912.11048}}.

\bibitem{Lasenby:2019prg}
R.~Lasenby, \emph{{Microwave cavity searches for low-frequency axion dark
  matter}},  \href{https://arxiv.org/abs/1912.11056}{{\ttfamily 1912.11056}}.

\bibitem{Lasenby:2019hfz}
R.~Lasenby, \emph{{Parametrics of electromagnetic searches for axion dark
  matter}},  \href{https://arxiv.org/abs/1912.11467}{{\ttfamily 1912.11467}}.

\bibitem{Horns2013a}
D.~Horns, J.~Jaeckel, A.~Lindner, A.~Lobanov, J.~Redondo and A.~Ringwald,
  \emph{{Searching for WISPy Cold Dark Matter with a Dish Antenna}},
  \href{https://doi.org/10.1088/1475-7516/2013/04/016}{\emph{JCAP} {\bfseries
  1304} (2013) 016}, [\href{https://arxiv.org/abs/1212.2970}{{\ttfamily
  1212.2970}}].

\bibitem{Baryakhtar:2018doz}
M.~Baryakhtar, J.~Huang and R.~Lasenby, \emph{{Axion and hidden photon dark
  matter detection with multilayer optical haloscopes}},
  \href{https://doi.org/10.1103/PhysRevD.98.035006}{\emph{Phys. Rev.}
  {\bfseries D98} (2018) 035006},
  [\href{https://arxiv.org/abs/1803.11455}{{\ttfamily 1803.11455}}].

\bibitem{Marsh:2018dlj}
D.~J. Marsh, K.-C. Fong, E.~W. Lentz, L.~Smejkal and M.~N. Ali, \emph{{Proposal
  to Detect Dark Matter using Axionic Topological Antiferromagnets}},
  \href{https://doi.org/10.1103/PhysRevLett.123.121601}{\emph{Phys. Rev. Lett.}
  {\bfseries 123} (2019) 121601},
  [\href{https://arxiv.org/abs/1807.08810}{{\ttfamily 1807.08810}}].

\bibitem{Hochberg:2016ajh}
Y.~Hochberg, T.~Lin and K.~M. Zurek, \emph{{Detecting Ultralight Bosonic Dark
  Matter via Absorption in Superconductors}},
  \href{https://doi.org/10.1103/PhysRevD.94.015019}{\emph{Phys. Rev.}
  {\bfseries D94} (2016) 015019},
  [\href{https://arxiv.org/abs/1604.06800}{{\ttfamily 1604.06800}}].

\bibitem{Schutz:2016tid}
K.~Schutz and K.~M. Zurek, \emph{{Detectability of Light Dark Matter with
  Superfluid Helium}},
  \href{https://doi.org/10.1103/PhysRevLett.117.121302}{\emph{Phys. Rev. Lett.}
  {\bfseries 117} (2016) 121302},
  [\href{https://arxiv.org/abs/1604.08206}{{\ttfamily 1604.08206}}].

\bibitem{Knapen:2016cue}
S.~Knapen, T.~Lin and K.~M. Zurek, \emph{{Light Dark Matter in Superfluid
  Helium: Detection with Multi-excitation Production}},
  \href{https://doi.org/10.1103/PhysRevD.95.056019}{\emph{Phys. Rev.}
  {\bfseries D95} (2017) 056019},
  [\href{https://arxiv.org/abs/1611.06228}{{\ttfamily 1611.06228}}].

\bibitem{Hochberg:2016sqx}
Y.~Hochberg, T.~Lin and K.~M. Zurek, \emph{{Absorption of light dark matter in
  semiconductors}},
  \href{https://doi.org/10.1103/PhysRevD.95.023013}{\emph{Phys. Rev.}
  {\bfseries D95} (2017) 023013},
  [\href{https://arxiv.org/abs/1608.01994}{{\ttfamily 1608.01994}}].

\bibitem{Knapen:2017ekk}
S.~Knapen, T.~Lin, M.~Pyle and K.~M. Zurek, \emph{{Detection of Light Dark
  Matter With Optical Phonons in Polar Materials}},
  \href{https://doi.org/10.1016/j.physletb.2018.08.064}{\emph{Phys. Lett.}
  {\bfseries B785} (2018) 386--390},
  [\href{https://arxiv.org/abs/1712.06598}{{\ttfamily 1712.06598}}].

\bibitem{Griffin:2018bjn}
S.~Griffin, S.~Knapen, T.~Lin and K.~M. Zurek, \emph{{Directional Detection of
  Light Dark Matter with Polar Materials}},
  \href{https://doi.org/10.1103/PhysRevD.98.115034}{\emph{Phys. Rev.}
  {\bfseries D98} (2018) 115034},
  [\href{https://arxiv.org/abs/1807.10291}{{\ttfamily 1807.10291}}].

\bibitem{Acanfora:2019con}
F.~Acanfora, A.~Esposito and A.~D. Polosa, \emph{{Sub-GeV Dark Matter in
  Superfluid He-4: an Effective Theory Approach}},
  \href{https://doi.org/10.1140/epjc/s10052-019-7057-0}{\emph{Eur. Phys. J. C}
  {\bfseries 79} (2019) 549},
  [\href{https://arxiv.org/abs/1902.02361}{{\ttfamily 1902.02361}}].

\bibitem{Trickle:2019ovy}
T.~Trickle, Z.~Zhang and K.~M. Zurek, \emph{{Detecting Light Dark Matter with
  Magnons}}, \href{https://doi.org/10.1103/PhysRevLett.124.201801}{\emph{Phys.
  Rev. Lett.} {\bfseries 124} (2020) 201801},
  [\href{https://arxiv.org/abs/1905.13744}{{\ttfamily 1905.13744}}].

\bibitem{Caputo:2019cyg}
A.~Caputo, A.~Esposito and A.~D. Polosa, \emph{{Sub-MeV Dark Matter and the
  Goldstone Modes of Superfluid Helium}},
  \href{https://doi.org/10.1103/PhysRevD.100.116007}{\emph{Phys. Rev. D}
  {\bfseries 100} (2019) 116007},
  [\href{https://arxiv.org/abs/1907.10635}{{\ttfamily 1907.10635}}].

\bibitem{Trickle:2019nya}
T.~Trickle, Z.~Zhang, K.~M. Zurek, K.~Inzani and S.~Griffin,
  \emph{{Multi-Channel Direct Detection of Light Dark Matter: Theoretical
  Framework}}, \href{https://doi.org/10.1007/JHEP03(2020)036}{\emph{JHEP}
  {\bfseries 03} (2020) 036},
  [\href{https://arxiv.org/abs/1910.08092}{{\ttfamily 1910.08092}}].

\bibitem{Griffin:2019mvc}
S.~M. Griffin, K.~Inzani, T.~Trickle, Z.~Zhang and K.~M. Zurek,
  \emph{{Multichannel direct detection of light dark matter: Target
  comparison}}, \href{https://doi.org/10.1103/PhysRevD.101.055004}{\emph{Phys.
  Rev. D} {\bfseries 101} (2020) 055004},
  [\href{https://arxiv.org/abs/1910.10716}{{\ttfamily 1910.10716}}].

\bibitem{Baym:2020uos}
G.~Baym, D.~Beck, J.~P. Filippini, C.~Pethick and J.~Shelton, \emph{{Searching
  for low mass dark matter via phonon creation in superfluid 4He}},
  \href{https://arxiv.org/abs/2005.08824}{{\ttfamily 2005.08824}}.

\bibitem{Martin2004a}
R.~M. Martin, \emph{Electronic Structure}.
\newblock Cambridge University Press, apr, 2004,
  \href{https://doi.org/10.1017/cbo9780511805769}{10.1017/cbo9780511805769}.

\bibitem{Kresse1993a}
G.~Kresse and J.~Hafner, \emph{{Ab initio molecular dynamics for liquid
  metals}}, \href{https://doi.org/10.1103/physrevb.47.558}{\emph{Physical
  Review B} {\bfseries 47} (jan, 1993) 558--561}.

\bibitem{Kresse1994a}
G.~Kresse and J.~Hafner, \emph{{Ab initio molecular-dynamics simulation of the
  liquid-metal{\textendash}amorphous-semiconductor transition in germanium}},
  \href{https://doi.org/10.1103/physrevb.49.14251}{\emph{Physical Review B}
  {\bfseries 49} (may, 1994) 14251--14269}.

\bibitem{Kresse1996a}
G.~Kresse and J.~Furthmuller, \emph{{Efficiency of ab-initio total energy
  calculations for metals and semiconductors using a plane-wave basis set}},
  \href{https://doi.org/10.1016/0927-0256(96)00008-0}{\emph{Computational
  Materials Science} {\bfseries 6} (jul, 1996) 15--50}.

\bibitem{Togo2015a}
A.~Togo and I.~Tanaka, \emph{First principles phonon calculations in materials
  science},
  \href{https://doi.org/10.1016/j.scriptamat.2015.07.021}{\emph{Scripta
  Materialia} {\bfseries 108} (Nov., 2015) 1--5}.

\bibitem{Toth-Lake}
S.~Toth and B.~Lake, \emph{{Linear spin wave theory for single-Q incommensurate
  magnetic structures}},  \href{https://arxiv.org/abs/1402.6069}{{\ttfamily
  1402.6069}}.

\bibitem{Colpa1978a}
J.~Colpa, \emph{Diagonalization of the quadratic boson hamiltonian},
  \href{https://doi.org/10.1016/0378-4371(78)90160-7}{\emph{Physica A:
  Statistical Mechanics and its Applications} {\bfseries 93} (sep, 1978)
  327--353}.

\bibitem{Saga}
V.~Cherepanov, I.~Kolokolov and V.~L'vov, \emph{{The saga of YIG: Spectra,
  thermodynamics, interaction and relaxation of magnons in a complex magnet}},
  \href{https://doi.org/10.1016/0370-1573(93)90107-O}{\emph{Physics Reports}
  {\bfseries 229} (jul, 1993) 81}.

\bibitem{Princep_2017}
A.~J. Princep, R.~A. Ewings, S.~Ward, S.~T{\'{o}}th, C.~Dubs, D.~Prabhakaran
  et~al., \emph{The full magnon spectrum of yttrium iron garnet},
  \href{https://doi.org/10.1038/s41535-017-0067-y}{\emph{npj Quantum Materials}
  {\bfseries 2} (nov, 2017) 63}.

\bibitem{Chigusa2020a}
S.~Chigusa, T.~Moroi and K.~Nakayama, \emph{Detecting light boson dark matter
  through conversion into magnon},
  \href{https://arxiv.org/abs/2001.10666}{{\ttfamily 2001.10666}}.

\bibitem{Cox2019a}
P.~Cox, T.~Melia and S.~Rajendran, \emph{Dark matter phonon coupling},
  \href{https://doi.org/10.1103/PhysRevD.100.055011}{\emph{Phys. Rev.}
  {\bfseries D100} (2019) 055011},
  [\href{https://arxiv.org/abs/1905.05575}{{\ttfamily 1905.05575}}].

\bibitem{Bhagavantam1939}
S.~Bhagavantam and T.~Venkatarayudu, \emph{Raman effect in relation to crystal
  structure}, \href{https://doi.org/10.1007/bf03046465}{\emph{Proceedings of
  the Indian Academy of Sciences - Section A} {\bfseries 9} (mar, 1939)
  224--258}.

\bibitem{PhysRev.132.1474}
A.~S. Barker, \emph{Infrared lattice vibrations and dielectric dispersion in
  corundum}, \href{https://doi.org/10.1103/PhysRev.132.1474}{\emph{Phys. Rev.}
  {\bfseries 132} (Nov, 1963) 1474--1481}.

\bibitem{PhysRevB.61.8187}
M.~Schubert, T.~E. Tiwald and C.~M. Herzinger, \emph{Infrared dielectric
  anisotropy and phonon modes of sapphire},
  \href{https://doi.org/10.1103/PhysRevB.61.8187}{\emph{Phys. Rev. B}
  {\bfseries 61} (Mar, 2000) 8187--8201}.

\bibitem{Tabuchi2015a}
Y.~Tabuchi, S.~Ishino, A.~Noguchi, T.~Ishikawa, R.~Yamazaki, K.~Usami et~al.,
  \emph{Quantum magnonics: magnon meets superconducting qubit},
  \href{https://arxiv.org/abs/1508.05290v1}{{\ttfamily 1508.05290v1}}.

\bibitem{Nielsen:1975hm}
H.~B. Nielsen and S.~Chadha, \emph{{On How to Count Goldstone Bosons}},
  \href{https://doi.org/10.1016/0550-3213(76)90025-0}{\emph{Nucl. Phys. B}
  {\bfseries 105} (1976) 445--453}.

\bibitem{Watanabe:2011ec}
H.~Watanabe and T.~Brauner, \emph{{On the number of Nambu-Goldstone bosons and
  its relation to charge densities}},
  \href{https://doi.org/10.1103/PhysRevD.84.125013}{\emph{Phys. Rev. D}
  {\bfseries 84} (2011) 125013},
  [\href{https://arxiv.org/abs/1109.6327}{{\ttfamily 1109.6327}}].

\bibitem{Watanabe:2012hr}
H.~Watanabe and H.~Murayama, \emph{{Unified Description of Nambu-Goldstone
  Bosons without Lorentz Invariance}},
  \href{https://doi.org/10.1103/PhysRevLett.108.251602}{\emph{Phys. Rev. Lett.}
  {\bfseries 108} (2012) 251602},
  [\href{https://arxiv.org/abs/1203.0609}{{\ttfamily 1203.0609}}].

\bibitem{Watanabe:2014fva}
H.~Watanabe and H.~Murayama, \emph{{Effective Lagrangian for Nonrelativistic
  Systems}}, \href{https://doi.org/10.1103/PhysRevX.4.031057}{\emph{Phys. Rev.
  X} {\bfseries 4} (2014) 031057},
  [\href{https://arxiv.org/abs/1402.7066}{{\ttfamily 1402.7066}}].

\bibitem{Kim_2019}
K.~Kim, S.~Y. Lim, J.-U. Lee, S.~Lee, T.~Y. Kim, K.~Park et~al.,
  \emph{Suppression of magnetic ordering in {XXZ}-type antiferromagnetic
  monolayer {NiPS}$_3$},
  \href{https://doi.org/10.1038/s41467-018-08284-6}{\emph{Nature
  Communications} {\bfseries 10} (jan, 2019) }.

\bibitem{Tanabashi:2018oca}
{\scshape Particle Data Group} collaboration, M.~Tanabashi et~al., \emph{Review
  of particle physics},
  \href{https://doi.org/10.1103/PhysRevD.98.030001}{\emph{Phys.Rev.D}
  {\bfseries 98} (2018) 030001}.

\bibitem{Visinelli2013a}
L.~Visinelli, \emph{Axion-electromagnetic waves},
  \href{https://doi.org/10.1142/S0217732313501629}{\emph{Mod. Phys. Lett.}
  {\bfseries A28} (2013) 1350162},
  [\href{https://arxiv.org/abs/1401.0709}{{\ttfamily 1401.0709}}].

\bibitem{Millar:2016cjp}
A.~J. Millar, G.~G. Raffelt, J.~Redondo and F.~D. Steffen, \emph{{Dielectric
  Haloscopes to Search for Axion Dark Matter: Theoretical Foundations}},
  \href{https://doi.org/10.1088/1475-7516/2017/01/061}{\emph{JCAP} {\bfseries
  01} (2017) 061}, [\href{https://arxiv.org/abs/1612.07057}{{\ttfamily
  1612.07057}}].

\bibitem{Ayala2014a}
A.~Ayala, I.~Dom{\'\i}nguez, M.~Giannotti, A.~Mirizzi and O.~Straniero,
  \emph{Revisiting the bound on axion-photon coupling from globular clusters},
  \href{https://doi.org/10.1103/PhysRevLett.113.191302}{\emph{Phys. Rev. Lett.}
  {\bfseries 113} (2014) 191302},
  [\href{https://arxiv.org/abs/1406.6053}{{\ttfamily 1406.6053}}].

\bibitem{MillerBertolami2014a}
M.~M. Miller~Bertolami, B.~E. Melendez, L.~G. Althaus and J.~Isern,
  \emph{Revisiting the axion bounds from the galactic white dwarf luminosity
  function}, \href{https://doi.org/10.1088/1475-7516/2014/10/069}{\emph{JCAP}
  {\bfseries 1410} (2014) 069},
  [\href{https://arxiv.org/abs/1406.7712}{{\ttfamily 1406.7712}}].

\bibitem{Wildes2015a}
A.~R. Wildes, V.~Simonet, E.~Ressouche, G.~J. McIntyre, M.~Avdeev, E.~Suard
  et~al., \emph{Magnetic structure of the quasi-two-dimensional antiferromagnet
  {NiPS$_3$}}, \href{https://doi.org/10.1103/physrevb.92.224408}{\emph{Physical
  Review B} {\bfseries 92} (dec, 2015) }.

\bibitem{LachanceQuirion2019a}
D.~Lachance-Quirion, Y.~Tabuchi, A.~Gloppe, K.~Usami and Y.~Nakamura,
  \emph{Hybrid quantum systems based on magnonics},
  \href{https://arxiv.org/abs/1902.03024}{{\ttfamily 1902.03024}}.

\end{thebibliography}\endgroup
\bibliographystyle{JHEP}

\end{document}